\documentclass[a4paper]{article}%
\usepackage{amsmath}
\usepackage{amsfonts}
\usepackage{amssymb}
\usepackage{graphicx}
\usepackage{cite}%
\providecommand{\U}[1]{\protect\rule{.1in}{.1in}}
\providecommand{\U}[1]{\protect\rule{.1in}{.1in}}
\providecommand{\U}[1]{\protect\rule{.1in}{.1in}}
\providecommand{\U}[1]{\protect\rule{.1in}{.1in}}
\providecommand{\U}[1]{\protect\rule{.1in}{.1in}}
\providecommand{\U}[1]{\protect\rule{.1in}{.1in}}
\providecommand{\U}[1]{\protect\rule{.1in}{.1in}}
\providecommand{\U}[1]{\protect\rule{.1in}{.1in}}
\providecommand{\U}[1]{\protect\rule{.1in}{.1in}}
\providecommand{\U}[1]{\protect\rule{.1in}{.1in}}
\providecommand{\U}[1]{\protect\rule{.1in}{.1in}}
\providecommand{\U}[1]{\protect\rule{.1in}{.1in}}
\providecommand{\U}[1]{\protect\rule{.1in}{.1in}}
\providecommand{\U}[1]{\protect\rule{.1in}{.1in}}
\providecommand{\U}[1]{\protect\rule{.1in}{.1in}}
\providecommand{\U}[1]{\protect\rule{.1in}{.1in}}
\providecommand{\U}[1]{\protect\rule{.1in}{.1in}}
\providecommand{\U}[1]{\protect\rule{.1in}{.1in}}
\providecommand{\U}[1]{\protect\rule{.1in}{.1in}}
\providecommand{\U}[1]{\protect\rule{.1in}{.1in}}
\providecommand{\U}[1]{\protect\rule{.1in}{.1in}}
\providecommand{\U}[1]{\protect\rule{.1in}{.1in}}
\providecommand{\U}[1]{\protect\rule{.1in}{.1in}}
\providecommand{\U}[1]{\protect\rule{.1in}{.1in}}
\providecommand{\U}[1]{\protect\rule{.1in}{.1in}}
\providecommand{\U}[1]{\protect\rule{.1in}{.1in}}

\begin{document}

\title{Interferometry with independent Bose-Einstein condensates: parity as an
EPR/Bell quantum variable}
\author{F. Lalo\"{e}$^{a}$ and W. J. Mullin$^{b}$\\$^{a}$Laboratoire Kastler Brossel, ENS, UPMC, CNRS ; 24 rue Lhomond, 75005
Paris, France\\$^{b}$Department of Physics, University of Massachusetts, Amherst,
Massachusetts 01003 USA}
\maketitle

\begin{abstract}
When independent Bose-Einstein condensates (BEC), described quantum
mechanically by Fock (number) states, are sent into interferometers, the
measurement of the output port at which the particles are detected provides a
binary measurement, with two possible results $\pm1$. With two interferometers
and two BEC's, the parity (product of all results obtained at each
interferometer) has all the features of an Einstein-Podolsky-Rosen quantity,
with perfect correlations predicted by quantum mechanics when the settings
(phase shifts of the interferometers) are the same.\ When they are different,
significant violations of Bell inequalities are obtained.\ These violations do
not tend to zero when the number $N$ of particles increases, and can therefore
be obtained with arbitrarily large systems, but a condition is that all
particles should be detected. We discuss the general experimental requirements
for observing such effects, the necessary detection of all particles in
correlation, the role of the pixels of the CCD detectors, and that of the
alignments of the interferometers in terms of matching of the wave fronts of
the sources in the detection regions.

Another scheme involving three interferometers and three BEC's is discussed;
it leads to Greenberger-Horne-Zeilinger (GHZ) sign contradictions, as in the
usual GHZ\ case with three particles, but for an arbitrarily large number of
them.\ Finally, generalizations of the Hardy impossibilities to an arbitrarily
large number of particles are introduced. BEC's provide a large versality for
observing violations of local realism in a variety of experimental arrangements.

\end{abstract}


\bigskip
The original Einstein-Podolsky-Rosen (EPR) argument \cite{EPR} considers a
system of two microscopic particles that are correlated; assuming that various
types of measurements are performed on this system in remote locations, and
using local realism, it shows that the system contains more \textquotedblleft
elements of reality\textquotedblright\ than those contained in quantum
mechanics.\ Bohr gave a refutation of the argument \cite{Bohr} by pointing out
that intrinsic physical properties should not be attributed to microscopic
systems, independently of their measurement apparatuses; in his view of
quantum mechanics (often called \textquotedblleft orthodox\textquotedblright),
the notion of reality introduced by EPR\ is inappropriate.\ Later, Bell
extended the EPR\ argument and used inequalities to show that local realism
and quantum mechanics may sometimes lead to contradictory predictions
\cite{Bell}.\ Using pairs of correlated photons emitted in a cascade, Clauser
et al. \cite{FC} checked that, even in this case, the results of quantum
mechanics are correct; other experiments leading to the same conclusion were
performed by Fry et al. \cite{Fry}, Aspect et al.\ \cite{Aspect}, and many
others.\ The body of all results is now such that it is generally agreed that
violations of the Bell inequalities do occur in Nature, even if experiments
are never perfect and if \textquotedblleft loopholes\textquotedblright\ (such
as sample bias \cite{Pearle, CH, CS}) can still be invoked in principle.\ All
these experiments were made with a small number of particles, generally a pair
of photons, so that Bohr's point of view directly applies to them.

In this article, as in \cite{FL} we consider systems made of an arbitrarily
large number of particles, and study some of their variables that can lead to
an EPR argument and Bell inequalities. Mermin \cite{Mermin} has also
considered a physical system made of many particles with spins, assuming that
the initial state is a so called GHZ state \cite{GHZ-1, GHZ-2}; another
many-particle quantum state has been studied by Drummond \cite{Drummond}%
.\ Nevertheless, it turns out that considering a double Fock state (DFS)\ with
spins, instead of these states, sheds interesting new light on the
Einstein-Bohr debate.\ The reason is that, in this case, the EPR\ elements of
reality can be macroscopic, for instance the total angular momentum (or
magnetization) contained in a large region of space; even if not measured,
such macroscopic quantities presumably possess physical reality, which gives
even more strength to the EPR argument.\ Moreover, one can no longer invoke
the huge difference of scales between the measured properties and the
measurement apparatuses, and Bohr's refutation becomes less plausible.

Double Fock states with spins also lead to violations of the Bell inequalities
\cite{PRL, LM}, so that they are appropriate for experimental tests of quantum
violations of local realism.\ A difficulty, nevertheless, is that the
violations require that all $N$ spins be measured in $N$ different regions of
space, which may be very difficult experimentally if $N$ exceeds $2$ or $3$;
with present experimental techniques, the schemes discussed in \cite{PRL, LM}
are therefore probably more thought experiments than realistic possibilities.
Here we come closer to experiments by studying schemes involving only
individual position measurement of the particles, without any necessity of
accurate localization.

With Bose condensed gases of metastable helium atoms, micro-channel plates
indeed allow one to detect atoms one by one \cite{Saubamea, Robert}.\ The
first idea that then comes to mind is to consider the interference pattern
created by a DFS, a situation that has been investigated theoretically by
several authors \cite{Java, WCW, CGNZ, PS, Dragan}, and observed
experimentally \cite{WK}.\ The quantum effects occurring in the detection of
the fringes have been studied in \cite{Polkovnikov-1, Polkovnikov-2}, in
particular the quantum fluctuations of the fringe amplitude; see also
\cite{DBRP} for a discussion of fringes observed with three condensates, in
correlation with populations oscillations. But, for obtaining quantum
violations of Bell type inequalities, continuous position measurements are not
necessarily optimal; it is more natural to consider measurement apparatuses
with a dichotomic result, such as interferometers with two outputs, as in
\cite{CD, SC}. Experimentally, laser atomic fluorescence may be used to
determine at which output of an interferometer atoms are found, without
requiring a very accurate localization of their position; in fact, since this
measurement process has a small effect on the measured quantity (the position
of the atom in one of the arms), one obtains in this way a quantum
non-demolition scheme where many fluorescence cycles can be used to ensure
good efficiency.

Quantum effects taking place in measurements performed with interferometers
with 2 input and 2 output ports (Mach-Zhender interferometers) have been
studied by several authors; refs \cite{HB, Dowling, DBB} discuss the effect of
quantum noise on an accurate measurement of phase, and compare the feeding of
interferometers with various quantum states; refs \cite{PS-1, PS-2, PS-3} give
a detailed treatment of the Heisenberg limit as well as of the role of the
Fisher information and of the Cramer-Rao lower bound in this problem.\ But, to
our knowledge, none of these studies leads to violations of Bell inequalities
and local realism.\ Here, we will consider interferometers with 4 input ports
and 4 output ports, in which a DFS is used to feed two of the inputs (the
others receive vacuum), and 4 detectors count the particles at the 4 output
ports - see Fig. \ref{fig1}; we will also consider a similar 6 input-6 output
case.\ This can be seen as a generalization of the work described by Yurke and
Stoler in refs.\ \cite{YS-1, YS-2}, and also to some extent of the
Rarity-Tapster experiment \cite{Franson, RT} (even if, in that case, the two
photons were not emitted by independent sources).

Another aspect of the present work is to address the question raised long ago
by Anderson \cite{Anderson} in the context of a thought experiment and, more
recently, by Leggett and Sols \cite{LS, Leggett}: \textquotedblleft Do
superfluids that have never seen each other have a well-defined relative
phase?\textquotedblright.\ \ A positive answer occurs in the usual view: when
spontaneous symmetry breaking takes place at the Bose- Einstein transition,
each condensate acquires a well-defined phase, though with a completely random
value.\ However, in quantum mechanics, the Bose-Einstein condensates of
superfluids are naturally described by Fock states, for which the phase of the
system is completely undetermined, in contradiction with this view.
Nevertheless, the authors of refs. \cite{Java, WCW, CGNZ, PS} and \cite{CD,
MKL} have shown how repeated quantum measurements of the relative phase of two
Fock states make a well-defined value emerge spontaneously, with a random
value. This seems to remove the contradiction; considering that the relative
phase appears under the effect of spontaneous symmetry breaking, as soon as
the BEC's are formed, or later, under the effect of measurements, then appears
as a matter of personal preference.

But a closer examination of the problem shows that this is not always true
\cite{PRL, Polkovnikov-2, LS, Leggett}: situations do exist where the two
points of view are not equivalent, and where the predictions of quantum
mechanics for an ensemble of measurements are at variance with those obtained
from a classical average over a phase.\ This is not so surprising after all:
the idea of a pre-existing phase is very similar to the notion of \ an
EPR\ \textquotedblleft element of reality\textquotedblright\ - for a double
Fock state, the relative phase is nothing but what is often called a
\textquotedblleft hidden variable\textquotedblright.\ The tools offered by the
Bell theorem are therefore appropriate to exhibit contradictions between the
notion of a pre--existing phase and the predictions of quantum
mechanics.\ Indeed, we will obtain violations of the BCHSH inequalities
\cite{BCHSH}, new GHZ contradictions \cite{GHZ-1, GHZ-2} as well as Hardy
impossibilities \cite{H-1, H-2}. Fock-state condensates appear as remarkably
versatile, able to create violations that usually require elaborate entangled
wave functions, and produce new $N$-body violations.

A preliminary short version of this work has been published in \cite{PRL-2}.
The present article gives more details and focusses on some issues that will
inevitably appear in the planning of an experiment, such as the effect of
non-perfect detection efficiency, losses, or the geometry of the wavefronts in
the region of the detectors.\ In \S \ \ref{quantum} we basically use the same
method as in \cite{PRL-2} (unitary transformations of creation operators),
following refs \cite{YS-1} and \cite{YS-2}, but include the treatment of
losses; in \S \ \ref{more elaborate}, we develop a more elaborate theory of
many-particle detection and high order correlation signals, performing a
calculation in $3N$ configuration space, and including a treatment of the
geometrical effects of wavefronts in the detection regions (this section may
be skipped by the reader who is not interested in experimental
considerations); finally, \S \ \ref{EPR} applies these calculations to three
situations: BCHSH inequality violations with two sources, GHZ contradictions
with three sources, and Hardy contradictions. Appendix I summarizes some
useful technical calculations; appendix II extends the calculations to initial
states other than the double Fock state (\ref{1}), in particular coherent and
phase states.

\section{Quantum calculation}

\label{quantum}
\begin{figure}[h]
\centering \includegraphics[width=3in]{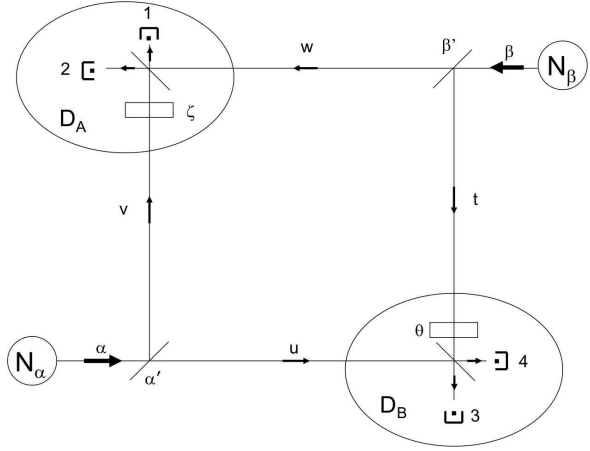}\caption{
Two Fock states, with populations $N_{\alpha}$ and $N_{\beta}$, enter beam
splitters, and are then made to interfere in two different regions of space
$D_{A}$ and $D_{B}$, with detectors 1 and 2 in the former, 3 and 4 in the
latter.\ The number of particles $m_{j}$ in each of the channels $j=1,2,3,4$
are counted.}%
\label{fig1}%
\end{figure}

We first calculate the prediction of quantum mechanics for the
experiment that is shown schematically in Fig. \ref{fig1}.\ \ Each of two
Bose-Einstein condensates, described by Fock states with populations
$N_{\alpha}$ and $N_{\beta}$, crosses a beam splitter; both are then made to
interfere at two other beam splitters, sitting in remote regions of space
$D_{A}$ and $D_{B}$. There, two operators, Alice and Bob, count the number of
particles that emerge from outputs 1 and 2 for Alice, outputs 3 and 4 for
Bob.\ By convention, channels 1 and 3 are ascribed a result $\eta=+1$,
channels 2 and 4 a result $\eta=-1$. We call $m_{j}$ the number of particles
that are detected at output $j$ (with $j=1$, $2$, $3$, $4$), $m_{A}%
=m_{1}+m_{2}$ the total number of particles detected by Alice, $m_{B}%
=m_{3}+m_{4}$ the number of particles detected by Bob, and $M=m_{A}+m_{B}$ the
total number of detected particles.\ From the series of results that they
obtain in each run of the experiment, both operators can calculate various
functions $\mathcal{A(}\eta_{1},..\eta_{m_{A}})$ and $\mathcal{B(}\eta
_{m_{A}+1},..\eta_{M})$ of their results; we will focus on the case where they
choose the parity, given by the product of all their $\eta$'s: $\mathcal{A}%
=(-1)^{m_{2}}$ for Alice, $\mathcal{B}=(-1)^{m_{4}}$ for Bob; for a discussion
of other possible choices, see \cite{LM}.

We now calculate the probability of any sequence of results with the same
approach as in \cite{PRL-2}.\ This provides correct results if one assumes
that the experiment is perfect; a more elaborate approach is necessary to
study the effects of experimental imperfections, and will be given in
\S \ \ref{more elaborate}.

\subsection{Probabilities of the various results}

\label{PVR}We consider spinless particles and we assume that the initial state
is:
\begin{equation}
\left\vert \Phi_{0}\right\rangle =\left\vert N_{\alpha},N_{\beta}\right\rangle
\equiv~\frac{1}{\sqrt{N_{\alpha}!N_{\beta}!}}~\left[  \left(  a_{\alpha
}\right)  ^{\dagger}\right]  ^{N_{\alpha}}\left[  \left(  a_{\beta}\right)
^{\dagger}\right]  ^{N_{\beta}}\left\vert 0\right\rangle \label{1}%
\end{equation}
where $\left\vert 0\right\rangle $ is the vacuum state; single particle state
$\alpha$ corresponds to that populated by the first source, $\beta$ to that
populated by the second source. The destruction operators $a_{1}\cdots a_{4}$
of the output modes can be written in terms of those of the modes at the
sources $a_{\alpha},$ $a_{\beta},$ $a_{\alpha^{\prime}}$ and $a_{\beta
^{\prime}}$ (including the vacuum input modes, $a_{\alpha^{\prime}}$ and
$a_{\beta^{\prime}}$, which are included to maintain unitarity) by tracing
back from the detectors to the sources, providing a phase shift of $\pi/2$ at
each reflection and $\zeta$ or $\theta$ at the shifters, and a $1/\sqrt{2}$
for normalization at each beam splitter. Thus we find:
\begin{equation}
\left(
\begin{array}
[c]{l}%
a_{1}\\
a_{2}\\
a_{3}\\
a_{4}%
\end{array}
\right)  =\frac{1}{2}\left(
\begin{array}
[c]{llll}%
ie^{i\zeta} & e^{i\zeta} & i & -1\\
-e^{i\zeta} & ie^{i\zeta} & 1 & i\\
i & -1 & ie^{i\theta} & e^{i\theta}\\
1 & i & -e^{i\theta} & e^{i\theta}%
\end{array}
\right)  \left(
\begin{array}
[c]{l}%
a_{\alpha}\\
a_{\alpha^{\prime}}\\
a_{\beta}\\
a_{\beta^{\prime}}%
\end{array}
\right)
\end{equation}

Since $a_{\alpha^{\prime}}$ and $a_{\beta^{\prime}}$ do not contribute we can
write simply:%

\begin{equation}%
\begin{array}
[c]{l}%
a_{1}=\frac{1}{2}\left[  ie^{i\zeta}a_{\alpha}+ia_{\beta}\right] \\
a_{2}=\frac{1}{2}\left[  -e^{i\zeta}a_{\alpha}+a_{\beta}\right] \\
a_{3}=\frac{1}{2}\left[  ia_{\alpha}+ie^{i\theta}a_{\beta}\right] \\
a_{4}=\frac{1}{2}\left[  a_{\alpha}-e^{i\theta}a_{\beta}\right]
\end{array}
\label{awm}%
\end{equation}
In short, we write these expressions as:
\begin{equation}
a_{j}=v_{j\alpha}a_{\alpha}+v_{j\beta}a_{\beta}. \label{ai}%
\end{equation}
We suppose that Alice finds $m_{1}$ positive results and $m_{2}$ negative
results for a total of $m_{A}$ measurements; Bob finds $m_{3}$ positive and
$m_{4}$ negative results in his $m_{B}$ total measurements. The quantum
probability of this series of results is the squared modulus of the scalar
product of state $\left\vert \Phi_{0}\right\rangle $ by the state associated
with the measurement:%
\begin{equation}
\mathcal{P(}m_{1},m_{2},m_{3},m_{4})=\frac{1}{m_{1}!m_{2}!m_{3}!m_{4}%
!}~\left\vert \left\langle 0\right\vert \left(  a_{1}\right)  ^{m_{1}}%
\cdots\left(  a_{4}\right)  ^{m_{4}}\left\vert N_{\alpha},N_{\beta
}\right\rangle \right\vert ^{2} \label{proba}%
\end{equation}
where the matrix element is non-zero only if:%
\begin{equation}
m_{1}+m_{2}+m_{3}+m_{4}=N_{\alpha}+N_{\beta}=N \label{sum-m}%
\end{equation}
We can calculate this matrix element by substituting (\ref{1}) and (\ref{ai})
and expanding in binomial series:%
\begin{equation}%
\begin{array}
[c]{l}%
\displaystyle\left\langle 0\right\vert \left(  a_{1}\right)  ^{m_{1}}%
\cdots\left(  a_{1}\right)  ^{m_{4}}\left\vert N_{\alpha},N_{\beta
}\right\rangle =\frac{1}{\sqrt{N_{\alpha}!N_{\beta}!}}\left\langle
0\right\vert \prod_{j=1}^{4}\left(  v_{j\alpha}a_{\alpha}+v_{j\beta}a_{\beta
}\right)  ^{m_{j}}\left(  a_{\alpha}^{\dagger}\right)  ^{N_{\alpha}}\left(
a_{\beta}^{\dagger}\right)  ^{N_{\beta}}\left\vert 0\right\rangle \\
\multicolumn{1}{c}{\displaystyle=\frac{1}{\sqrt{N_{\alpha}!N_{\beta}!}}%
\sum_{p_{\alpha1}=0}^{m_{1}}\frac{m_{1}!}{p_{\alpha1}!p_{\beta1}!}\left(
v_{1\alpha}\right)  ^{p_{\alpha1}}\left(  v_{1\beta}\right)  ^{p_{\beta1}}%
...}\\
\multicolumn{1}{r}{\displaystyle..\sum_{p_{\alpha4}=0}^{m_{4}}\frac{m_{4}%
!}{p_{\alpha4}!p_{\beta4}!}\left(  v_{4\alpha}\right)  ^{p_{\alpha4}}\left(
v_{4\beta}\right)  ^{p_{\beta4}}~\left\langle 0\right\vert \left(  a_{\alpha
}\right)  ^{p_{\alpha1}+\cdots+p_{\alpha4}}\left(  a_{\beta}\right)
^{p_{\beta1}+\cdots+p_{\beta4}m_{i}}\left(  a_{\alpha}^{\dagger}\right)
^{N_{\alpha}}\left(  a_{\beta}^{\dagger}\right)  ^{N_{\beta}}\left\vert
0\right\rangle }%
\end{array}
\label{CWM}%
\end{equation}
where $p_{\beta j}=m_{j}-p_{\alpha j}$ for any $j$. The matrix element at the
end of this expression is:%
\begin{equation}
N_{\alpha}!N_{\beta}!~\delta_{N_{\alpha},~p_{\alpha1}+\cdots+~p_{\alpha4}%
}~\delta_{N_{\beta},~p_{\beta1}+\cdots~+p_{\beta4}} \label{matrix element}%
\end{equation}
But by definition the sum of all $p$'s is equal to $m_{1}+m_{2}+m_{3}+m_{4}$
which, according to (\ref{sum-m}), is $N_{\alpha}+N_{\beta}$; the two
Kronecker delta's in (\ref{matrix element}) are therefore redundant.\ For the
matrix element to be non zero and equal to $N_{\alpha}!N_{\beta}!$, it is
sufficient that the difference between the sums $p_{\alpha1}+\cdots
+~p_{\alpha4}$ and $p_{\beta1}+\cdots~+p_{\beta4}$ be equal to $N_{\alpha
}-N_{\beta}$, a condition which we can express through the integral:
\begin{equation}
\int_{-\pi}^{\pi}\frac{d\mu}{2\pi}e^{i(N_{\beta}-N_{\alpha}+p_{\alpha1}%
+\cdots+~p_{\alpha4}-p_{\beta1}+\cdots~-p_{\beta4})\mu}=\delta_{N_{\alpha
}-N_{\beta},~p_{\alpha1}+\cdots+~p_{\alpha4}-p_{\beta1}-\cdots~-p_{\beta4}}
\label{delta}%
\end{equation}
When this is inserted into (\ref{CWM}), in the second line every $v_{j\alpha}$
becomes $v_{j\alpha}e^{i\mu}$, every $v_{j\beta}$ becomes $v_{j\beta}e^{-i\mu
}$, and we can redo the sums and write the probability amplitude as:
\begin{equation}
\sqrt{N_{\alpha}!N_{\beta}!}\int_{-\pi}^{\pi}\frac{d\mu}{2\pi}e^{^{i\left(
N_{\beta}-N_{\alpha}\right)  \mu}}\prod_{j=1}^{4}\left(  v_{j\alpha}e^{i\mu
}+v_{j\beta}e^{-i\mu}\right)  ^{m_{j}} \label{ampl}%
\end{equation}

Thus the probability is:%
\begin{equation}
\mathcal{P(}m_{1},m_{2},m_{3},m_{4})=\frac{N_{\alpha}!N_{\beta}!}{m_{1}%
!m_{2}!m_{3}!m_{4}!}\int_{-\pi}^{\pi}\frac{d\mu}{2\pi}\int_{\pi}^{\pi}%
\frac{d\mu^{\prime}}{2\pi}~~e^{^{i\left(  N_{\beta}-N_{\alpha}\right)  \left(
\mu-\mu^{\prime}\right)  }}\prod_{j=1}^{4}\left[  \Omega_{j}^{\ast}%
(\mu^{\prime})\Omega_{j}(\mu)\right]  ^{m_{i}} \label{prob-2}%
\end{equation}
with:%
\begin{equation}
\Omega_{j}(\mu)=v_{j\alpha}e^{i\mu}+v_{j\beta}e^{-i\mu} \label{omega}%
\end{equation}
Each of the factors $\Omega_{j}^{\ast}(\mu^{\prime})\Omega_{j}(\mu)$ can now
be simplified according to:%
\begin{equation}
\Omega_{j}^{\ast}(\mu^{\prime})\Omega_{j}(\mu)=\left\vert v_{j\alpha
}\right\vert ^{2}e^{i\left(  \mu-\mu^{\prime}\right)  }+\left\vert v_{j\beta
}\right\vert ^{2}e^{i\left(  \mu^{\prime}-\mu\right)  }+v_{j\alpha}^{\ast
}v_{j\beta}~e^{-i\left(  \mu+\mu^{\prime}\right)  }+v_{j\alpha}v_{j\beta
}^{\ast}~e^{i\left(  \mu+\mu^{\prime}\right)  } \label{produit-omega}%
\end{equation}
which, when (\ref{awm}) is inserted, gives:%
\[
\frac{1}{2}\left[  \cos\left(  \mu-\mu^{\prime}\right)  \pm\cos\left(
\zeta+\mu+\mu^{\prime}\right)  \right]  ~~~~~\text{or}~~~~~\frac{1}{2}\left[
\cos\left(  \mu-\mu^{\prime}\right)  \pm\cos\left(  -\theta+\mu+\mu^{\prime
}\right)  \right]
\]
depending on the value of $j$. Now, if we define:%
\begin{equation}%
\begin{array}
[c]{l}%
\lambda=\mu+\mu^{\prime}\\
\Lambda=\mu-\mu^{\prime}%
\end{array}
\label{Lambda}%
\end{equation}
we finally obtain:%
\begin{equation}%
\begin{array}
[c]{l}%
\displaystyle\mathcal{P(}m_{1},m_{2},m_{3},m_{4})=\frac{N_{\alpha}!N_{\beta}%
!}{m_{1}!m_{2}!m_{3}!m_{4}!}~2^{-N}\int_{-\pi}^{\pi}\frac{d\lambda}{2\pi}%
\int_{-\pi}^{\pi}\frac{d\Lambda}{2\pi}\cos\left[  (N_{\beta}-N_{\alpha
})\Lambda\right] \\
\multicolumn{1}{r}{\displaystyle\times\left[  \cos\Lambda+\cos\left(
\zeta+\lambda\right)  \right]  ^{m_{1}}\left[  \cos\Lambda-\cos\left(
\zeta+\lambda\right)  \right]  ^{m_{2}}\left[  \cos\Lambda+\cos\left(
\theta-\lambda\right)  \right]  ^{m_{3}}\left[  \cos\Lambda-\cos\left(
\theta-\lambda\right)  \right]  ^{m_{4}}}%
\end{array}
\label{19}%
\end{equation}
where we have used $\Lambda$ parity to reduce $e^{i\left(  N_{\beta}%
-N_{\alpha}\right)  \Lambda}$ to a cosine.

\subsection{Effects of particle losses}

We now study cases where losses occur in the experiment; some particles
emitted by the sources are missed by the detectors sitting at the four output
ports.\ The total number of particles they detect is $M$, with $M\leq N$; an
analogous situation was already considered in the context of spin measurements
\cite{PRL, LM}.\ We first focus on losses taking place near the sources of
particles, then on those in the detection regions.

\subsubsection{Losses at the sources}

\label{losses-sources}
\begin{figure}[h]
\centering \includegraphics[width=3in]{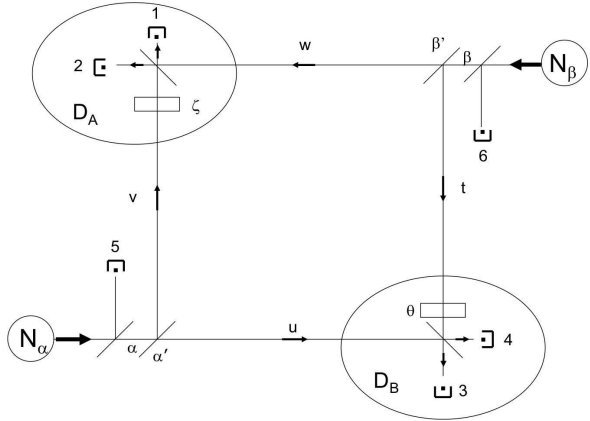} \caption{The
experiment is the same as in figure \ref{fig1}, but now we assume that two
beam splitters are inserted between the two sources and the inputs of the
interferometer. Then, the total number of particles $M$ measured at the output
of the interferometer may be less than $N$.}%
\label{fig2}%
\end{figure}

As a first simple model for treating losses, we consider
the experimental configuration shown in Fig.\ \ref{fig2}, where additional
beam splitters divert some particles before they reach the input of the
interferometer.\ If $T$ and $R$ are the transmission and the reflection
coefficients of the additional beam splitters, with:%
\begin{equation}
R+T=1 \label{6a-bis}%
\end{equation}
the unitary transformations become:%
\begin{equation}%
\begin{array}
[c]{l}%
a_{1}=\frac{i\sqrt{T}}{2}\left[  e^{i\zeta}a_{\alpha}+\alpha_{\beta}\right] \\
a_{2}=\frac{\sqrt{T}}{2}\left[  -e^{i\zeta}a_{\alpha}+\alpha_{\beta}\right] \\
a_{3}=\frac{i\sqrt{T}}{2}\left[  a_{\alpha}+e^{i\theta}a_{\beta}\right] \\
a_{4}=\frac{\sqrt{T}}{2}\left[  a_{\alpha}-e^{i\theta}a_{\beta}\right] \\
a_{5}=i\sqrt{R}\left[  a_{\alpha}\right] \\
a_{6}=i\sqrt{R}\left[  a_{\beta}\right]
\end{array}
\label{6a}%
\end{equation}

The probability amplitude associated with a series of results $m_{1}$,
...$m_{5}$, $m_{6}$ is now:%
\begin{equation}
\frac{\left\langle 0\right\vert \left(  a_{1}\right)  ^{m_{1}}\cdots\left(
a_{4}\right)  ^{m_{4}}\left(  a_{5}\right)  ^{m_{5}}\left(  a_{6}\right)
^{m_{6}}\left\vert N_{\alpha},N_{\beta}\right\rangle }{\sqrt{m_{1}%
!....m_{5}!~m_{6}!}} \label{s1}%
\end{equation}
or, taking into account the last two equations (\ref{6a}):%
\begin{equation}
R^{^{\left(  m_{5}+m_{6}\right)  /2}}\sqrt{\frac{N_{\alpha}!}{\left(
N_{\alpha}-m_{5}\right)  !~m_{5}!}\times\frac{N_{\beta}!}{\left(  N_{\beta
}-m_{6}\right)  !~m_{6}!}}\frac{\left\langle 0\right\vert \left(
a_{1}\right)  ^{m_{1}}\cdots\left(  a_{4}\right)  ^{m_{4}}\left\vert
N_{\alpha}-m_{5},N_{\beta}-m_{6}\right\rangle }{\sqrt{m_{1}!...m_{4}!}}
\label{t1}%
\end{equation}
The fraction on the right of this expression can be obtained from the
calculations of \S \ \ref{PVR}, by just replacing $N_{\alpha}$ by $N_{\alpha
}-m_{5}$, $N_{\beta}$ by $N_{\beta}-m_{6}$ in (\ref{ampl}).\ With this
substitution, the numerical factor in front of that expression combines with
that of (\ref{t1}) to give a prefactor:%
\begin{equation}
R^{^{\left(  m_{5}+m_{6}\right)  /2}}\sqrt{\frac{N_{\alpha}!N_{\beta}!}%
{m_{5}!m_{6}!}} \label{pref}%
\end{equation}

The next step is to sum the probabilities over $m_{5}$ and $m_{6}$, keeping
the four $m_{1}$, ..$m_{4}$ constant; we vary $m_{5}$ and $m_{6}$ with a
constant sum $N-M$, where $M$ is defined as:%
\begin{equation}
M=m_{1}+m_{2}+m_{3}+m_{4}\leq N \label{def-M}%
\end{equation}
The $\cos\left[  \left(  N_{\beta}-N_{\alpha}\right)  \Lambda\right]  $ inside
the integral of (\ref{19}) arose from an exponential $e^{i\left[  \left(
N_{\beta}-N_{\alpha}\right)  \Lambda\right]  }$, which now becomes:%
\begin{equation}
e^{i\left(  N_{\beta}-N_{\alpha}+m_{5}-m_{6}\right)  \Lambda} \label{expo}%
\end{equation}
so that the summation over $m_{5}$ and $m_{6}$ reconstructs a power of a
binomial:%
\begin{equation}
\frac{1}{\left(  N-M\right)  !}\left[  e^{i\Lambda}+e^{-i\Lambda}\right]
^{N-M}=\frac{2^{(N-M)}}{\left(  N-M\right)  !}\left[  \cos\Lambda\right]
^{N-M} \label{binom}%
\end{equation}
When the powers of $R$ and $T$ are included as well as the factors $2^{-M}$
and $2^{(N-M)}$, equation (\ref{19}) is now replaced by:%
\begin{equation}%
\begin{array}
[c]{l}%
\displaystyle\mathcal{P}_{M}\mathcal{(}m_{1},m_{2},m_{3},m_{4})=\frac
{N_{\alpha}!N_{\beta}!}{m_{1}!m_{2}!m_{3}!m_{4}!}2^{N-2M}\frac{T^{M}~R^{N-M}%
~}{\left(  N-M\right)  !}\int_{-\pi}^{\pi}\frac{d\lambda}{2\pi}\int_{-\pi
}^{\pi}\frac{d\Lambda}{2\pi}\cos\left[  (N_{\beta}-N_{\alpha})\Lambda\right]
~\left[  \cos\Lambda\right]  ^{N-M}\\
\multicolumn{1}{r}{\displaystyle\times\left[  \cos\Lambda+\cos\left(
\zeta+\lambda\right)  \right]  ^{m_{1}}\left[  \cos\Lambda-\cos\left(
\zeta+\lambda\right)  \right]  ^{m_{2}}}\\
\multicolumn{1}{r}{\displaystyle\times\left[  \cos\Lambda+\cos\left(
\theta-\lambda\right)  \right]  ^{m_{3}}\left[  \cos\Lambda-\cos\left(
\theta-\lambda\right)  \right]  ^{m_{4}}}%
\end{array}
\label{Proba}%
\end{equation}
where $M$ is defined in (\ref{def-M}).\ This result is similar to (\ref{19}),
but includes a power of $\cos\Lambda$ inside the integral, which we will
discuss in \S \ \ref{discussion}. We note that this power of $\cos\Lambda$
introduces exactly the same factor as that already obtained in \cite{PRL}, in
the context of spin condensates and particles missed in transverse spin measurements.

If $T=1$ and $R=0$, expression (\ref{Proba}) vanishes unless $M$ has its
maximal value $M=N$; then expression (\ref{19}) is recovered, as expected. If
$R$ and $T$ have intermediate values, $M$ has a probability distribution
including any value less than $N$, with of course smaller values favored when
$T$ is small and $R$ large.

\subsubsection{Losses at the detectors}

\label{losses-detectors}
\begin{figure}[h]
\centering \includegraphics[width=3in]{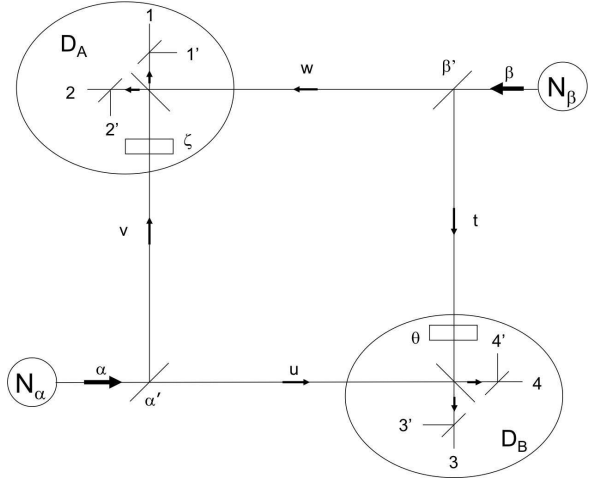} \caption{The
experiment is the same as that in figure 2, but now 4 beam splitters are
inserted just before the 4 particle detectors, which sit in channels $1$%
,$2$,$3$,$4$; the other channels, $1^{\prime}$, $2^{\prime}$, $3^{\prime}$,
$4^{\prime}$ contain no detector.\ This provides a model for calculating the
effect of limited quantum efficiencies of the detectors.}%
\label{fig2-2}%
\end{figure}

Instead of inserting additional beam splitters just
after the sources, we can put them just before the detectors, as in Fig.
\ref{fig2-2}; this provides a model for losses corresponding to imperfect
detectors with quantum efficiencies less than $100\%$.

\ Instead of 6 destruction operators as in (\ref{6a}), we now have 8; each
operator $a_{j}$ ($j=1,2,3,4)$ corresponding to one of the 4 the detectors is
now associated with a second operator $a_{j}^{\prime}$ corresponding to the
other output port with no detector.\ For instance, for $j=1$, one has:%
\begin{equation}
a_{1}=\frac{i\sqrt{T}}{2}\left[  e^{i\zeta}a_{\alpha}+\alpha_{\beta}\right]
\text{ \ \ \ \ \ \ ; \ \ \ \ \ \ \ \ }a_{1}^{\prime}=\frac{-\sqrt{R}}%
{2}\left[  e^{i\zeta}a_{\alpha}+\alpha_{\beta}\right]  \label{a-prime}%
\end{equation}
and similar results for $j=2$,$3$,$4$.\ The calculation of the probability
associated with results $m_{1}$,..$m_{4}$ (with their sum equal to $M$) and
$m_{1}^{\prime}$, ..$m_{4}^{\prime}$ (with their sum equal to $N-M$) is very
similar to that of \S \ \ref{PVR}; formula (\ref{proba}) becomes:%
\[
\mathcal{P(}m_{1},..m_{4};m_{1}^{\prime},..m_{4}^{\prime})=\frac{1}%
{m_{1}!..!m_{4}!\times m_{1}^{\prime}!..!m_{4}^{\prime}!}~\left\vert
\left\langle 0\right\vert \left(  a_{1}\right)  ^{m_{1}}\cdots\left(
a_{4}\right)  ^{m_{4}}\times\left(  a_{1}^{\prime}\right)  ^{m_{1}^{\prime}%
}\cdots\left(  a_{4}^{\prime}\right)  ^{m_{4}^{\prime}}\left\vert N_{\alpha
},N_{\beta}\right\rangle \right\vert ^{2}%
\]
Since $a_{j}$ and $a_{j}^{\prime}$ are almost the same operator (they just
differ by a coefficient), the result is still given by the right hand side of
(\ref{19}), with the following changes:

(i) each $m_{j}$ is now replaced by the sum $m_{j}+m_{j}^{\prime}$

(ii) $m_{1}!..!m_{4}!$ in the denominator is multiplied by $m_{1}^{\prime
}!..!m_{4}^{\prime}!$

(ii) a factor $T^{M}\times R^{N-M}$ appears in front of the expression.

Now, we consider the observed results $m_{1}$,..$m_{4}$ as fixed, and add the
probabilities associated with all possible non-observed values $m_{1}^{\prime
}$, ..$m_{4}^{\prime}$;\ this amounts to distributing $N-M$ unobserved
particles in any possible way among all output channels without detectors. The
summation is made in two steps:

(i) summations over $m_{1}^{\prime}$ and $m_{2}^{\prime}$ at constant sum
$m_{1}^{\prime}+m_{2}^{\prime}=m_{A}^{\prime}$, and over $m_{3}^{\prime}$ and
$m_{4}^{\prime}$ at constant sum $m_{3}^{\prime}+m_{4}^{\prime}=m_{B}^{\prime
}$

(ii) summation over $m_{A}^{\prime}$ and $m_{B}^{\prime}$ at constant sum
$m_{A}^{\prime}+m_{B}^{\prime}=N-M$

The first summation provides:%
\begin{equation}
\sum_{m_{1}^{\prime}+m_{2}^{\prime}=m_{A}^{\prime}}\frac{1}{m_{1}^{\prime
}!m_{2}^{\prime}!}\left[  \cos\Lambda+\cos\left(  \zeta+\lambda\right)
\right]  ^{m_{1}^{\prime}}\left[  \cos\Lambda-\cos\left(  \zeta+\lambda
\right)  \right]  ^{m_{2}^{\prime}}=\frac{1}{m_{A}^{\prime}!}\left[
2\cos\Lambda\right]  ^{m_{A}^{\prime}} \label{sum1}%
\end{equation}
and similarly for the summation over $m_{3}^{\prime}$ and $m_{4}^{\prime}$.
Then the last summation provides:%
\begin{equation}
\sum_{m_{A}^{\prime}+m_{B}^{\prime}=N-M}\frac{1}{m_{A}^{\prime}!m_{B}^{\prime
}!}\left[  2\cos\Lambda\right]  ^{m_{A}^{\prime}}\left[  2\cos\Lambda\right]
^{m_{B}^{\prime}}=\frac{1}{N-M!}\left[  4\cos\Lambda\right]  ^{N-M}
\label{sum2}%
\end{equation}
At the end, as in \S \ \ref{losses-sources}, we see that each unobserved
particle introduces a factor $\cos\Lambda$ so that, in the integral giving the
probability, a factor $\left[  \cos\Lambda\right]  ^{N-M}$ appears; actually,
we end up with an expression that is again exactly (\ref{Proba}).\ The
presence of this factor inside the integral seems to be a robust property of
the effects of imperfect measurements (for brevity, we do not prove the
generality of this statement, for instance by studying the effect of
additional beam splitters that are inserted elsewhere, for instance in other
parts of the interferometer).

\subsection{Discussion}

\label{discussion}The discussion of the physical content of equation
(\ref{19}) and (\ref{Proba}) is somewhat similar to that for spin measurements
\cite{PRL}.\ One difference is that, here, we consider that the total number
of measurements $m_{A}=m_{1}+m_{2}$ made by Alice, as well as the total number
of measurements $m_{B}=m_{3}+m_{4}$ made by Bob, are left to fluctuate with a
constant sum $M$; the particles emitted by the sources may localize in any of
the four detection regions.\ With spins, the numbers of detections depends on
the number of spin apparatuses used by Alice and Bob, so that it was more
natural to assume that $m_{A}$ and $m_{B}$ are fixed.\ This changes the
normalization and the probabilities, but not the dependence on the
experimental parameters, which is given by (\ref{19}) and (\ref{Proba}). A
discussion of the normalization integrals is given in the Appendix.

\subsubsection{Effects of $N_{\alpha}$, $N_{\beta}$ and of the number of
measurements}

\label{par}If the numbers of particles in the sources are different$\ $%
($N_{\alpha}\neq N_{\beta}$), a term in $\cos\left[  (N_{\beta}-N_{\alpha
})\Lambda\right]  $ appears in both equations; let us assume for simplicity
that all $N$ particles are measured $(M=N)$, so that equation (\ref{19})
applies, and for instance that $N_{\alpha}>N_{\beta}$. Then, in the product of
factors inside the integral, only some terms can provide a non-zero
contribution after integration over $\Lambda$; we must choose at least
$N_{\alpha}-N_{\beta}$ factors contributing through $\cos\left(
\Lambda\right)  $, and thus at most $N-(N_{\alpha}-N_{\beta})=2N_{\beta}$
factors contributing through the $\theta$ dependent terms.\ Therefore
$2N_{\beta}$ is the maximum number of particles providing results that depends
on the settings of the interferometer; all the others have equal probabilities
$1/2$, whatever the phase shift is.\ This is physically understandable, since
$(N_{\alpha}-N_{\beta})$ particles from the first source unmatched particles
from the other, and can thus not contribute to an interference effect. All
particles can contribute coherently to the interference only if $N_{\alpha
}=N_{\beta}$.

If the numbers of particles in the sources are equal ($N_{\alpha}=N_{\beta}$),
the sources are optimal; equation (\ref{Proba}) contains the effect of missing
some particles in the measurements.\ If the number of experiments $M$ is much
less than a very large $N$, because $\cos\Lambda^{N-M}$ peaks up
sharply\footnote{Here we take the point of view where the $\Lambda$
integration domain is between $-\pi/2$ and $+\pi/2$; otherwise, we should also
take into account a peak around $\Lambda=\pi$.} at $\Lambda=0$, the result
simplifies into:%

\begin{align}
&  \mathcal{P}_{M}\mathcal{(}m_{1},m_{2},m_{3},m_{4})\sim\frac{1}{m_{1}%
!m_{2}!m_{3}!m_{4}!}\int_{-\pi}^{\pi}\frac{d\lambda}{2\pi}~~\left[
1+\cos\left(  \zeta+\lambda\right)  \right]  ^{m_{1}}\left[  1-\cos\left(
\zeta+\lambda\right)  \right]  ^{m_{2}}\nonumber\\
&  \times\left[  1+\cos\left(  \theta-\lambda\right)  \right]  ^{m_{3}}\left[
1-\cos\left(  \theta-\lambda\right)  \right]  ^{m_{4}} \label{22}%
\end{align}
We then recover \textquotedblleft classical\textquotedblright\ results,
similar to those of refs.\ \cite{CD} or \cite{FL-2}.\ Suppose that we
introduce a classical phase $\lambda$ and calculate classically the
interference effects at both beam splitters.\ This leads to intensities
proportional to $\left[  1+\cos\left(  \zeta+\lambda\right)  \right]  $ and
$\left[  1-\cos\left(  \zeta+\lambda\right)  \right]  $ on both sides of the
interferometer in $D_{A}$, and similar results for $D_{B}$.\ Now, if we assume
that each particle reaching the beam splitter has crossing and reflecting
probabilities that are proportional to these intensities (we treat each of
these individual processes as independent), and if we consider that this
classical phase is completely unknown, an average over $2\pi$ then
reconstructs exactly (\ref{22}).\ In this case, the classical image of a
pre-existing phase leads to predictions that are the same as those of quantum
mechanics; this phase will take a completely random value for each realization
of the experiment, with for instance no way to force it to take related values
in two successive runs.\ All this fits well within the concept of the Anderson
phase, originating from spontaneous symmetry breaking at the phase transition
(Bose-Einstein condensation): at this transition point, the quantum system
\textquotedblleft chooses\textquotedblright\ a phase, which takes a completely
random value, and then plays the role of a classical variable (in the limit of
very large systems).

On the other hand, if $N-M$ vanishes, the peaking effect of $\cos\Lambda
^{N-M}$ does not occur anymore, $\Lambda$ can take values close to $\pi/2$ ,
so that the terms in the product inside the integral are no longer necessarily
positive; an interpretation in terms of classical probabilities then becomes
impossible.\ In these cases, the phase does not behave as a semi-classical
variable, but retains a strong quantum character; the variable $\Lambda$
controls the amount of quantum effects.\ It is therefore natural to call
$\Lambda$ the \textquotedblleft quantum angle\textquotedblright\ and $\lambda$
the \textquotedblleft classical phase\textquotedblright.\

One could object that, if expression (\ref{19}) contains negative factors,
this does not prove that the same probabilities $\mathcal{P}$ can not be
obtained with another mathematical expression without negative
probabilities.\ To show that this is indeed impossible, we have to resort to a
more general theorem, the Bell/BCHSH theorem, which proves it in a completely
general way; this is what we do in \S \ \ref{EPR}.

\subsubsection{Perfect correlations}

\label{perfect}We now show that, if $N_{\alpha}$ and $N_{\beta}$ are equal and
if the number of measurements is maximal ($M=N$), when Alice and Bob choose
opposite\footnote{Usually, perfect correlations are obtained when the two
settings are the same, not opposite.\ But, with the geometry shown in figure
\ref{fig1}, $\zeta$ introduces a phase delay of source $\alpha$ with respect
to source $\beta$, while $\theta$ does the opposite by delaying source $\beta$
with respect to source $\alpha$. Therefore, the dephasing effects of the two
delays are the same in both regions $D_{A}$ and $D_{B}$ when $\theta=-\zeta$.}
phase shifts ($\theta=-\zeta$) they always measure the same parity.\ In this
case, the integrand of (\ref{19}) becomes:%
\begin{equation}
\left[  \cos\Lambda+\cos\left(  \lambda+\zeta\right)  \right]  ^{m_{1}+m_{3}%
}\left[  \cos\Lambda-\cos\left(  \lambda+\zeta\right)  \right]  ^{m_{2+}m_{4}}
\label{integrand}%
\end{equation}
which can also be written as:%
\begin{equation}
2^{N}\left[  \sin(\lambda^{\prime}+\frac{\zeta}{2})\right]  ^{m_{1}+m_{3}%
}\left[  \sin(\lambda^{\prime\prime}-\frac{\zeta}{2})\right]  ^{m_{1}+m_{3}%
}\left[  \cos(\lambda^{\prime}+\frac{\zeta}{2})\right]  ^{m_{2}+m_{4}}\left[
\cos(\lambda^{\prime\prime}-\frac{\zeta}{2})\right]  ^{m_{2}+m_{4}}
\label{integrand-2}%
\end{equation}
with the following change of integration variables\footnote{Using the
periodicity of the integrand, one can give to both integration variables
$\lambda^{\prime}$ and $\lambda^{\prime\prime}$ a range $\left[  -\pi
,+\pi\right]  $; this doubles the integration domain, but this doubling is
cancelled by a factor $1/2$ introduced by the Jacobian.}:%
\begin{equation}%
\begin{array}
[c]{cc}%
\lambda^{\prime}=\frac{\lambda+\Lambda}{2} & \lambda^{\prime\prime}%
=\frac{\Lambda-\lambda}{2}%
\end{array}
\label{lambdas}%
\end{equation}
If, for instance, $m_{1}+m_{3}$ is odd, instead of $\lambda^{\prime}$ one can
take $(\lambda^{\prime}+\zeta/2)$ as an integration variable, and one can see
that the integral vanishes because its periodicity - the same is true of
course for the $\lambda^{\prime\prime}$ integration, which also
vanishes.\ Similarly, if $m_{2}+m_{4}$ is odd, one can take $(\lambda^{\prime
}-\frac{\zeta}{2})$ and $(\lambda^{\prime\prime}-\frac{\zeta}{2})$ as
integration variables, and the result vanishes again.\ Finally, the
probability is non-zero only if both $m_{1}+m_{3}$ and $m_{2}+m_{4}$ are even;
the conclusion is that Alice and Bob always observe the same parity for their
results.\ This perfect correlation is useful for applying the EPR\ reasoning
to parities.

\section{A more elaborate calculation}

\label{more elaborate}The advantage of measuring the positions of particles
after a beam splitter, that is interference effects providing dichotomic
results, is that one has a device that is close to quantum non-demolition
experiments.\ With a resonant laser, one can make an atom fluoresce and emit
many photons, without transferring the atom from one arm of the interferometer
to the other.\ This is clearly important in experiments where, as we have
seen, all the atoms must be detected to obtain quantum non-local effects.\ On
the other hand, it is well know experimentally that a difficulty with
interferometry is the alignment of the devices in order to obtain an almost
perfect matching of the wave front.\ We discuss this problem now.

A general assumption behind the calculation of \S \ \ref{quantum} is that, in
each region of space (for instance at the inputs, or at the 4 outputs), only
one mode of the field is populated (only one $a^{\dagger}$ operator is
introduced per region).\ The advantage of this approach is its simplicity, but
it nevertheless eludes some interesting questions.\ For instance, suppose that
the energies of the particles emerging from each source differ by some
arbitrarily small quantity; after crossing all beam splitters, they would
reach the detection regions in two orthogonal modes, so that the probability
of presence would be the sum of the corresponding probabilities, without any
interference term.\ On the other hand, all interesting effects obtained in
\cite{PRL-2} are precisely interference effects arising because, in each
detection region, there is no way to tell which source emitted the detected
particles.\ Does this mean that these effects disappear as soon as the sources
are not strictly identical, so that the quantum interferences will never be
observable in practice?

To answer this kind of question, here we will develop a more detailed theory
of the detection of many particles in coincidence, somewhat similar to
Glauber's theory of photon coincidences \cite{Glauber}; nevertheless, while in
that theory only the initial value of the n-th time derivative was calculated,
here we study the whole time dependence of the correlation function. Our
result will be that, provided the interferometers and detectors are properly
aligned, it is the detection process that restores the interesting quantum
effects, even if the sources do not emit perfectly identical wavefronts in the detection regions, as was assumed in the calculations of
\S \ \ref{quantum}.\ As a consequence,
the reader who is not interested in experimental limitations may skip this
section and proceed directly to \S \ \ref{EPR}.

In this section we change the definition of the single particle states:
$\alpha$ now corresponds to a state for which the wave function originates
from the first source, is split into two beams when reaching the first beam
splitter, and into two beams again when it reaches the beam splitters
associated with the regions of measurement $D_{A}$ and $D_{B}$; \ the same is
true for state $\beta$.\ In this point of view, all the propagation in the
interferometers is already included in the states.\ We note in passing that
the evolution associated with the beam splitters is unitary; the states
$\alpha$ and $\beta$ therefore remain perfectly orthogonal, even if they
overlap in some regions of space. Having changed the definition of the single
particle states, we keep (\ref{1}) to define the N particle state of the system.

\subsection{Pixels as independent detectors}

We model the detectors sitting after the beam splitters by assuming that they
are the juxtaposition of a large number $\mathcal{Q}$ of independent pixels,
which we treat as independent detectors.\ This does not mean that the
positions of the impact of all particles are necessarily registered in the
experiment; our calculations still apply if, for instance, only the total
number of impact in each channel is recorded.\ The only thing we assume is
that the detection of particles in different points leads to orthogonal states
of some part of the apparatus (or the environment), so that we can add the
probabilities of the events corresponding to different orthogonal states;
whether or not the information differentiating these states is recorded in
practice does not matter.

If the number of pixels $\mathcal{Q}$ is very large, the probability of
detection of two bosons at the same pixel is negligible.\ If we note
$N=N_{\alpha}+N_{\beta}$, this probability is bounded\footnote{We add the
probabilities of non-exclusive events, whith provides an upper bound of the
real probability of double detection.} by:%
\begin{equation}
\frac{1}{\mathcal{Q}}+\frac{2}{\mathcal{Q}}+....+\frac{N}{\mathcal{Q}}%
=\frac{N(N+1)}{2\mathcal{Q}} \label{2}%
\end{equation}
so that we will assume that:%
\begin{equation}
\mathcal{Q}\gg N^{2} \label{3}%
\end{equation}
Moreover, we consider events where all $N$ particles are detected (the
probabilities of events where some particles are missed can be obtained from
the probabilities of these events in a second step, as in \cite{PRL}).

\subsection{Flux of probability at the pixels in a stationary state}

\label{flux}Each pixel $j$ is considered as defining a region of space
$\Delta_{j}$ in which the particles are converted into a macroscopic electric
current, as in a photomultiplier.\ The particles enter this region through the
front surface $S_{j}$ of the pixel; all particles crossing $S_{j}$ disappear
in a conversion process that is assumed to have 100\% efficiency.\ What we
need, then, is to calculate the flux of particles entering the front surfaces
of the pixels.\ It is convenient to reason in the $3N$ dimension configuration
space, in which the hyper-volume associated with the $N$ different pixels is:%
\begin{equation}
V_{N}=\Delta_{1}\otimes\Delta_{2}\otimes\Delta_{3}....\otimes\Delta_{N}
\label{a-1}%
\end{equation}
which has an front surface given by:%
\begin{equation}
S_{N}=S_{1}\otimes\Delta_{2}\otimes\Delta_{3}....\otimes\Delta_{N}%
~+~\Delta_{1}\otimes S_{2}\otimes\Delta_{3}....\otimes\Delta_{N}%
+~..+~~\Delta_{1}\otimes\Delta_{2}\otimes\Delta_{3}....\otimes S_{N}
\label{a-1-s}%
\end{equation}
The density of probability in this space is defined in terms of the boson
field operator $\Psi(\mathbf{r})$ as:%
\begin{equation}
\rho_{N}(\mathbf{r}_{1},\mathbf{r}_{2},...,\mathbf{r}_{N})=\Psi^{\dagger
}(\mathbf{r}_{1})\Psi(\mathbf{r}_{1})~\Psi^{\dagger}(\mathbf{r}_{2}%
)\Psi(\mathbf{r}_{2})~...~\Psi^{\dagger}(\mathbf{r}_{N})\Psi(\mathbf{r}_{N})
\label{a-2}%
\end{equation}
where we assume that all $\mathbf{r}_{j}$'s are different (all pixels are
disjoint).\ The components of the $3N$ dimension current operator
$\mathbf{J}_{N}$ are:%
\begin{equation}
\mathbf{J}_{N}=\left\{
\begin{array}
[c]{l}%
\frac{\hbar}{2mi}\left[  \Psi^{\dagger}(\mathbf{r}_{1})\mathbf{\nabla}%
\Psi(\mathbf{r}_{1})-\mathbf{\nabla}\Psi^{\dagger}(\mathbf{r}_{1}%
)\Psi(\mathbf{r}_{1})\right]  ~~\Psi^{\dagger}(\mathbf{r}_{2})\Psi
(\mathbf{r}_{2})~...\Psi^{\dagger}(\mathbf{r}_{N})\Psi(\mathbf{r}_{N})+\\
+\frac{\hbar}{2mi}\Psi^{\dagger}(\mathbf{r}_{1})\Psi(\mathbf{r}_{1})~\left[
\Psi^{\dagger}(\mathbf{r}_{2})\mathbf{\nabla}\Psi(\mathbf{r}_{2}%
)-\mathbf{\nabla}\Psi^{\dagger}(\mathbf{r}_{2})\Psi(\mathbf{r}_{2})\right]
...\Psi^{\dagger}(\mathbf{r}_{N})\Psi(\mathbf{r}_{N})+\\
+...\\
+\frac{\hbar}{2mi}\Psi^{\dagger}(\mathbf{r}_{1})\Psi(\mathbf{r}_{1}%
)~\Psi^{\dagger}(\mathbf{r}_{2})\Psi(\mathbf{r}_{2})~...\left[  \Psi^{\dagger
}(\mathbf{r}_{N})\mathbf{\nabla}\Psi(\mathbf{r}_{N})-\mathbf{\nabla}%
\Psi^{\dagger}(\mathbf{r}_{N})\Psi(\mathbf{r}_{N})\right]
\end{array}
\right.  \label{a-4}%
\end{equation}
In the Heisenberg picture, the quantum operator $\rho_{N}$ obeys the evolution
equation:%
\begin{equation}
\frac{d}{dt}\rho_{N}(\mathbf{r}_{1},\mathbf{r}_{2},...,\mathbf{r}%
_{N};t)+\mathbf{\nabla}_{N}\cdot\mathbf{J}_{N}=0 \label{a-3}%
\end{equation}
where $\mathbf{\nabla}_{N}$ is the $N$ dimensional divergence.\ The flux of
probability entering the $3N$ dimension volume $S_{N}$ it then:%

\begin{equation}%
\begin{array}
[c]{l}%
\mathcal{F}(\Delta_{1},\Delta_{2},...\Delta_{N})~=~<\Phi_{0}\mid F\left(
\Delta_{1}\right)  \times G\left(  \Delta_{2}\right)  \times....G\left(
\Delta_{N}\right)  \mid\Phi_{0}>\\
\multicolumn{1}{r}{+~<\Phi_{0}\mid G\left(  \Delta_{1}\right)  \times F\left(
\Delta_{2}\right)  \times....G\left(  \Delta_{N}\right)  \mid\Phi_{0}>+...}\\
\multicolumn{1}{r}{+~<\Phi_{0}\mid G\left(  \Delta_{1}\right)  \times G\left(
\Delta_{2}\right)  \times....F\left(  \Delta_{N}\right)  \mid\Phi_{0}>}%
\end{array}
\label{4}%
\end{equation}
where $F(\Delta_{j})$ is the operator defined as a flux surface integral
associated to pixel $j$:
\begin{equation}
F\left(  \Delta_{j}\right)  =\frac{\hbar}{2mi}~\int_{S_{j}}d^{2}%
\mathbf{s\cdot}~\left[  \Psi^{\dagger}(\mathbf{r}^{^{\prime}})\mathbf{\nabla
}\Psi(\mathbf{r}^{^{\prime}})-\mathbf{\nabla}\Psi^{\dagger}(\mathbf{r}%
^{^{\prime}})\Psi(\mathbf{r}^{^{\prime}})\right]  \label{5}%
\end{equation}
($d^{2}\mathbf{s}$ the differential vector perpendicular to the surface) and
where $G(\Delta_{j})$ is a volume integral associated to the same pixel:%
\begin{equation}
G\left(  \Delta_{j}\right)  =\int_{\Delta_{j}}d^{3}r^{\prime}~\Psi^{\dagger
}(\mathbf{r}^{^{\prime}})\Psi(\mathbf{r}^{^{\prime}}) \label{5-bis}%
\end{equation}
The value of $\mathcal{F}(\Delta_{1},\Delta_{2},...\Delta_{N})$ provides the
time derivative of the probability of detection at all selected pixels, which
we calculate in \S \ \ref{probab}.

Now, because the various pixels do not overlap, the field operators commute
and we can push all $\Psi^{\dagger}$'s to the left, all $\Psi$'s to the right;
then we expand these operators on a basis that has $u_{\alpha}$ and $u_{\beta
}$ as its two first vectors:%
\begin{equation}
\Psi(\mathbf{r})=u_{\alpha}(\mathbf{r})~a_{\alpha}~+u_{\beta}(\mathbf{r}%
)~a_{\beta}+...~ \label{6}%
\end{equation}
The end of the expansion, noted $....$, corresponds to the components of
$\Psi(\mathbf{r})$ on other modes that must be added to modes $\alpha$ and
$\beta$ to form a complete orthogonal basis in the space of states of one
single particle; it is easy to see that they give vanishing contributions to
the average in state $\mid\Phi_{0}>$.\ The structure of any term in (\ref{4})
then becomes (for the sake of simplicity, we just write the first term):%
\begin{equation}%
\begin{array}
[c]{l}%
<\Phi_{0}\mid\mathcal{O}_{a,a\dagger}~~\frac{\hbar}{2mi}\left\{  \int
_{\Delta_{1}}\left[  u_{\alpha}^{\ast}(\mathbf{r}_{1}^{\prime})a_{\alpha
}^{\dagger}+u_{\beta}^{\ast}(\mathbf{r}_{1}^{\prime})a_{\beta}^{\dagger
}\right]  ~d^{2}\mathbf{s}_{1}\cdot\left[  \mathbf{\nabla}u_{\alpha
}(\mathbf{r}_{1}^{\prime})a_{\alpha}+\mathbf{\nabla}u_{\beta}(\mathbf{r}%
_{1}^{\prime})a_{\beta}\right]  -\text{c.c.}\right\}  \times\\
\multicolumn{1}{r}{\times%
{\displaystyle\prod\limits_{j=2}^{N}}
\left[  u_{\alpha}^{\ast}(\mathbf{r}_{j}^{\prime})a_{\alpha}^{\dagger
}+u_{\beta}^{\ast}(\mathbf{r}_{j}^{\prime})a_{\beta}^{\dagger}\right]  \left[
u_{\alpha}(\mathbf{r}_{j}^{\prime})a_{\alpha}+u_{\beta}(\mathbf{r}_{j}%
^{\prime})a_{\beta}\right]  \mid\Phi_{0}>}%
\end{array}
\label{7}%
\end{equation}
where c.c. means complex conjugate and where $\mathcal{O}_{a,a\dagger}$ is the
normal ordering operator that puts all the creation operators $a_{\alpha
,\beta}^{\dagger}$ to the left of all annihilation operators $a_{\alpha,\beta
}$. Each term of the product inside this matrix element contains a product of
operators that give, either zero, or always the same matrix element
$N_{\alpha}!N_{\beta}!$.\ For obtaining a non-zero value, two conditions are necessary:

(i) the number of $a_{\alpha}^{\dagger}$'s should be equal to that of
$a_{\alpha}$'s

(ii) the number of $a_{\alpha}$'s, minus that of $a_{\beta}$'s, should be
equal to $N_{\alpha}-N_{\beta}$.

These conditions are fulfilled with the help of two integrals:%
\begin{equation}
\int_{-\pi}^{\pi}\frac{d\lambda}{2\pi}~~\int_{-\pi}^{\pi}\frac{d\Lambda}{2\pi
}~~e^{i(N_{\beta}-N_{\alpha})\Lambda} \label{8}%
\end{equation}
and by multiplying:

(i) every $u_{\alpha}(\mathbf{r}_{j}^{\prime})$ by $e^{i\lambda}$, and every
$u_{\alpha}^{\ast}(\mathbf{r}_{j}^{\prime})$ by $e^{-i\lambda}$ (without
changing the wave functions related to $\beta$)

(ii) then every $u_{\alpha}(\mathbf{r}_{j}^{\prime})$ by $e^{i\Lambda}$, and
every $u_{\beta}(\mathbf{r}_{j}^{\prime})$ by $e^{-i\Lambda}$ (without
touching the complex conjugate wave functions).

This provides:%

\begin{equation}%
\begin{array}
[c]{l}%
\mathcal{F\sim}\int_{-\pi}^{\pi}\frac{d\lambda}{2\pi}~~\int_{-\pi}^{\pi}%
\frac{d\Lambda}{2\pi}~~e^{i(N_{\beta}-N_{\alpha})\Lambda}\\
\frac{\hbar}{2mi}\int_{\Delta_{1}}d^{2}\mathbf{s}_{1}\cdot\left[  u_{\alpha
}^{\ast}(\mathbf{r}_{1}^{\prime})\mathbf{\nabla}u_{\alpha}(\mathbf{r}%
_{1}^{\prime})e^{i\Lambda}+u_{\beta}^{\ast}(\mathbf{r}_{1}^{\prime
})\mathbf{\nabla}u_{\beta}(\mathbf{r}_{1}^{\prime})e^{-i\Lambda}+u_{\alpha
}^{\ast}(\mathbf{r}_{1}^{\prime})\mathbf{\nabla}u_{\beta}(\mathbf{r}%
_{1}^{\prime})e^{-i(\lambda+\Lambda)}+u_{\beta}^{\ast}(\mathbf{r}_{1}^{\prime
})\mathbf{\nabla}u_{\alpha}(\mathbf{r}_{1}^{\prime})e^{i(\lambda+\Lambda
)}\right. \\
\multicolumn{1}{r}{\left.  -u_{\alpha}(\mathbf{r}_{1}^{\prime})\mathbf{\nabla
}u_{\alpha}^{\ast}(\mathbf{r}_{1}^{\prime})e^{i\Lambda}-u_{\beta}%
(\mathbf{r}_{1}^{\prime})\mathbf{\nabla}u_{\beta}^{\ast}(\mathbf{r}%
_{1}^{\prime})e^{-i\Lambda}-u_{\alpha}(\mathbf{r}_{1}^{\prime})\mathbf{\nabla
}u_{\beta}^{\ast}(\mathbf{r}_{1}^{\prime})e^{i(\lambda+\Lambda)}-u_{\beta
}(\mathbf{r}_{1}^{\prime})\mathbf{\nabla}u_{\alpha}^{\ast}(\mathbf{r}%
_{1}^{\prime})e^{-i(\lambda+\Lambda)}\right] }\\
\multicolumn{1}{r}{\times%
{\displaystyle\prod\limits_{j=2}^{N}}
\left[  u_{\alpha}^{\ast}(\mathbf{r}_{j}^{\prime})u_{\alpha}(\mathbf{r}%
_{j}^{\prime})e^{i\Lambda}+u_{\beta}^{\ast}(\mathbf{r}_{j}^{\prime})u_{\beta
}(\mathbf{r}_{j}^{\prime})e^{-i\Lambda}+u_{\alpha}^{\ast}(\mathbf{r}%
_{j}^{\prime})u_{\beta}(\mathbf{r}_{j}^{\prime})e^{-i(\lambda+\Lambda
)}+\text{c.c.}\right]  ~+\text{sim.}}%
\end{array}
\label{9}%
\end{equation}
where \textquotedblleft sim.\textquotedblright\ is for the $N-1$ similar terms
where the gradients occur for $j=2,3,..N$, instead of $j=1$.

Now we assume that the experiment is properly aligned so that the wavefronts
of the wave functions $\alpha$ and $\beta$ coincide in all detection regions;
then, in the gradients:%
\begin{equation}
\mathbf{\nabla}u_{\alpha}(\mathbf{r})=iu_{\alpha}(\mathbf{r})~\mathbf{k}%
_{\alpha}~\text{ \ \ \ \ \ \ ; \ \ \ \ \ \ \ }\mathbf{\nabla}u_{\beta
}(\mathbf{r})=iu_{\beta}(\mathbf{r})~\mathbf{k}_{\beta}~ \label{10-c}%
\end{equation}
the vectors $\mathbf{k}_{\alpha}$ and $\mathbf{k}_{\beta}$ are parallel;
actually, on each pixel~$j$ we take these two vectors as equal to the same
constant value $\mathbf{k}_{\Delta_{j}}$, assuming that the pixels are small
and that the wavelengths of the two wave functions are almost equal.\ Then
(\ref{9}) simplifies into:%

\begin{equation}%
\begin{array}
[c]{l}%
\mathcal{F\sim}\int_{-\pi}^{\pi}\frac{d\lambda}{2\pi}~~\int_{-\pi}^{\pi}%
\frac{d\Lambda}{2\pi}~~e^{i(N_{\beta}-N_{\alpha})\Lambda}\frac{\hbar}{2m}%
\int_{S_{1}}\mathbf{k}_{\Delta_{1}}\cdot d^{2}\mathbf{s}_{1}...\int
_{\Delta_{2}}d^{3}r_{2}^{\prime}...\int_{\Delta_{i}}d^{3}r_{i}^{\prime}~...\\
\multicolumn{1}{r}{%
{\displaystyle\prod\limits_{j=1}^{N}}
\left[  u_{\alpha}^{\ast}(\mathbf{r}_{j}^{\prime})u_{\alpha}(\mathbf{r}%
_{j}^{\prime})e^{i\Lambda}+u_{\beta}^{\ast}(\mathbf{r}_{j}^{\prime})u_{\beta
}(\mathbf{r}_{j}^{\prime})e^{-i\Lambda}+u_{\alpha}^{\ast}(\mathbf{r}%
_{j}^{\prime})u_{\beta}(\mathbf{r}_{j}^{\prime})e^{-i(\lambda+\Lambda
)}+u_{\beta}^{\ast}(\mathbf{r}_{j}^{\prime})u_{\alpha}(\mathbf{r}_{j}^{\prime
})e^{i(\lambda+\Lambda)}\right]  ~+\text{ sim.}}%
\end{array}
\label{10-d}%
\end{equation}
In this expression, the integrand in the $\lambda$ and $\Lambda$ integrals is
a product of $N$ factors corresponding to the individual pixels, but this does
not imply the absence of correlations (the integral of a product is not the
product of integrals).\ The probability flux $\mathcal{F}$ contains surface
integrals through the front surfaces of the pixels, as expected, but also
volume integrals in other pixels, which is less intuitive\footnote{Equation
(\ref{a-1-s}) shows that, in the definition of surface in $3N$ dimension
space, the $2$ dimension front surface of any pixel is associated with all
three dimensions of any other pixel, including its depth.\ These dimensions
play the role of transverse dimensions over which an integration has to be
performed to obtain the flux (similarly, in $3$ dimensions, the flux through a
surface perpendicular to $Oz$ contains an integration over the transverse
directions $Ox$ and $Oy$).}.\ As a consequence, for obtaining a non-zero
probability flux $\mathcal{F}$ in $3N$ dimensions, it is not sufficient to
have a non-zero three dimension probability flux through one (or several)
pixels; it is also necessary that some probability has already accumulated in
the other pixels.\ In other words, at the very moment where the wave functions
reach the front surface of the pixels, the first time derivative of the
probability density in $3N$ dimension remains zero, while only the $N$-th
order time derivative is non-zero (this will be seen more explicitly in
\S \ \ref{time}).\ This is analogous to the photon detection process with $N$
atoms in quantum optics, see for instance Glauber \cite{Glauber}.

\subsection{Time dependence}

\label{time}Consider an experiment where each source emits a wave packet in a
finite time.\ We assume that, in each wave packet, all the particles still
remain in the same quantum state, but that this state is now time dependent;
the state of the system is then still given by (\ref{1}), but with time
dependent states $\alpha$ and $\beta$, so that the creation operators are now
$\left(  a_{\alpha}\right)  ^{\dagger}(t)$ and $\left(  a_{\beta}\right)
^{\dagger}(t)$. All the calculation of the previous section remains valid, the
main difference being that the wave functions are time dependent: $u_{\alpha
}(\mathbf{r},t)$ and $u_{\beta}(\mathbf{r},t)$.

We must therefore take into account possible time dependences of the wave
fronts of the two wave functions, as well as those of their amplitudes and
phases.\ If, for instance, the interferometer is perfectly symmetric, and if
the two wave packets are emitted at the same time, they will reach the beam
splitters of the detection regions at the same time with wave fronts that will
perfectly overlap at the detectors; the amplitudes of the two wave functions
will always be the same.\ See for instance ref.\ \cite{Hagley, Simsarian} for
a discussion of the time evolution of the phase of Bose-Einstein condensates,
including the effects of the interactions within the condensate.\

If we consider separately each factor inside the $\lambda$ and $\Lambda$
integral of (\ref{10-d}), we come back to the usual three dimension space; two
different kinds of integrals then occur:%
\begin{equation}%
\begin{array}
[c]{l}%
f_{j}(t)=\frac{\hbar}{2m}\int_{S_{j}}\mathbf{k}_{\Delta_{j}}\cdot
d^{2}\mathbf{s}_{j}\left[  u_{\alpha}^{\ast}(\mathbf{r}_{j}^{\prime
},t)u_{\alpha}(\mathbf{r}_{j}^{\prime},t)e^{i\Lambda}+u_{\beta}^{\ast
}(\mathbf{r}_{j}^{\prime},t)u_{\beta}(\mathbf{r}_{j}^{\prime},t)e^{-i\Lambda
}\right. \\
\multicolumn{1}{r}{\left.  +u_{\alpha}^{\ast}(\mathbf{r}_{j}^{\prime
},t)u_{\beta}(\mathbf{r}_{j}^{\prime},t)e^{-i(\lambda+\Lambda)}+u_{\beta
}^{\ast}(\mathbf{r}_{j}^{\prime},t)u_{\alpha}(\mathbf{r}_{j}^{\prime
},t)e^{i(\lambda+\Lambda)}\right] }%
\end{array}
\label{10-e}%
\end{equation}
and:%
\begin{equation}%
\begin{array}
[c]{l}%
g_{j}(t)=\int_{\Delta_{i}}d^{3}r_{i}^{\prime}\left[  u_{\alpha}^{\ast
}(\mathbf{r}_{j}^{\prime},t)u_{\alpha}(\mathbf{r}_{j}^{\prime},t)e^{i\Lambda
}+u_{\beta}^{\ast}(\mathbf{r}_{j}^{\prime},t)u_{\beta}(\mathbf{r}_{j}^{\prime
},t)e^{-i\Lambda}+\right. \\
\multicolumn{1}{r}{\left.  u_{\alpha}^{\ast}(\mathbf{r}_{j}^{\prime
},t)u_{\beta}(\mathbf{r}_{j}^{\prime},t)e^{-i(\lambda+\Lambda)}+u_{\beta
}^{\ast}(\mathbf{r}_{j}^{\prime},t)u_{\alpha}(\mathbf{r}_{j}^{\prime
},t)e^{i(\lambda+\Lambda)}\right] }%
\end{array}
\label{10-f}%
\end{equation}
The conservation law in ordinary space implies that $f_{j}(t)$ is related to
the time derivative of $g_{j}(t)$: it gives the contribution of the front
surface of the pixel to the time variation of the accumulated probability
$g_{j}(t)$ in volume $\Delta_{i}$.\ The total time derivative of $g_{j}(t)$ is
given by:%
\begin{equation}
\frac{d}{dt}g(t)=f_{j}(t)-f_{j}^{-}(t) \label{10-f-bis}%
\end{equation}
where $f_{j}^{-}(t)$ is the flux of the three dimensional probability current
through the lateral and rear surfaces of volume $\Delta_{j}$; the first term
in the right hand side is the entering flux, the second term the out-going
flux, with a leak through the rear surface that begins to be non-zero as soon
as the wave functions have crossed the entire detection volume $\Delta_{j}$.
But we do not have to take into account this out-going flux of probability: we
assume that the detection process absorbs all bosons.\ {For instance, }once
they enter volume $\Delta_{j}$, {the atoms are ionized and the emitted
electron is amplified into an cascade process, as in a photomultiplier; the
detection probability accumulated over time does not decrease under the effect
of $f_{j}^{-}(t)$. Therefore we must ignore the second term in the rhs of
(\ref{10-f-bis}), and replace (\ref{10-f}) by the more appropriate definition
of $g_{j}(t)$:%
\begin{equation}
g_{j}(t)=\int_{0}^{t}dt^{\prime}~f_{j}(t^{\prime}) \label{10-g}%
\end{equation}
(we assume that time $t=0$ occurs just before the wave packets reach the
detectors).\ With this relation, we no longer have to manipulate two
independent functions $f$ and $g$; moreover, the value of $g_{j}(t)$ now
depends only of the values of wave functions on the front surface of the
detector, which is physically satisfying (while (\ref{10-f}) contains
contributions of the wave functions in all volume $\Delta_{j}$, an unphysical
result if this volume has a large depth). }

We have already assumed in (\ref{10-c}) that the wavefronts of the two waves
are parallel on every pixel; we moreover assume that $\mathbf{k}_{\Delta_{j}}$
is perpendicular to the surface of the pixel, and then call $\varphi
(\Delta_{j})$ their relative phase over this pixel, taking it as a constant
over the pixel and over time, during the propagation of the wave functions
(which is the case if the interferometer is symmetrical, as in the
figure).\ Moreover, we assume that the two wave functions $u_{\alpha}$ and
$u_{\beta}$ have the same square modulus $\left\vert u_{\Delta_{j}%
}(t)\right\vert ^{2}$ at this pixel, so that the interference contrast is
optimal (again, this is related to a proper alignment of the
interferometer).\ Then (\ref{10-e}) becomes:%
\begin{equation}
f_{j}(t)\simeq\frac{\hbar}{2m}S_{j}\left\vert \mathbf{k}_{\Delta_{j}%
}\right\vert \left\vert u_{\Delta_{j}}(t)\right\vert ^{2}\left\{  \cos
\Lambda+\cos\left[  \varphi(\Delta_{j})-\Lambda-\lambda\right]  \right\}
=\frac{d}{dt}p_{j}(t)\times\left\{  \cos\Lambda+\cos\left[  \varphi(\Delta
_{j})-\Lambda-\lambda\right]  \right\}  \label{10-k}%
\end{equation}
with:%
\begin{equation}
p_{j}(t)=\frac{\hbar}{2m}S_{j}\left\vert \mathbf{k}_{\Delta_{j}}\right\vert
\int_{0}^{t}dt^{\prime}\left\vert u_{\Delta_{j}}(t^{\prime})\right\vert ^{2}
\label{10-l}%
\end{equation}
where $S_{j}$ is the area of pixel $j$.

Finally, we assume that all the pixels are identical so that their detection
areas have the same value $S$. We then obtain the simplified expression:%
\begin{equation}
\mathcal{F(}t\mathcal{)\sim~}S^{N}\int_{-\pi}^{\pi}\frac{d\lambda}{2\pi}%
~~\int_{-\pi}^{\pi}\frac{d\Lambda}{2\pi}~~\cos\left[  (N_{\beta}-N_{\alpha
})\Lambda\right]  ~~\frac{\text{d}}{\text{dt}}%
{\displaystyle\prod\limits_{j=1}^{N}}
p_{j}(t)\left\{  \cos\Lambda+\cos\left[  \varphi(\Delta_{j})-\Lambda
-\lambda\right]  \right\}  ~ \label{11}%
\end{equation}
(we have used $\Lambda$ parity to replace the exponential in $(N_{\beta
}-N_{\alpha})\Lambda$ by a cosine, so that the reality of the expression is
more obvious) and the accumulated probability at time $t$ is:%
\begin{equation}
\mathcal{P(}t\mathcal{)\sim}\int_{-\pi}^{\pi}\frac{d\lambda}{2\pi}~~\int
_{-\pi}^{\pi}\frac{d\Lambda}{2\pi}~~\cos\left[  (N_{\beta}-N_{\alpha}%
)\Lambda\right]  ~\times%
{\displaystyle\prod\limits_{j=1}^{N}}
p_{j}(t)~\left\{  \cos\Lambda+\cos\left[  \varphi(\Delta_{j})-\Lambda
-\lambda\right]  \right\}  \label{11-bis}%
\end{equation}

We finally consider a situation where $m_{1}$ pixels belong to the first
detector, $m_{2}$ to the second, etc., each sitting in one detection region
after the last beam splitters.\ We assume that the front surface of the
detectors are parallel to the wave fronts, so that all phases differences
$\varphi(\Delta_{j})$ collapse into 4 values only, two (in region $D_{A}$)
containing the phase shift $\zeta$, two (in region $D_{B}$) containing the
phase shift $\theta$:%
\begin{equation}%
\begin{array}
[c]{l}%
\varphi_{A}-\zeta\text{ for the }m_{1}\text{ first measurements}\\
\varphi_{A}-\zeta+\pi\text{ for the next }m_{2}\text{ measurements}\\
\varphi_{B}+\theta\text{ for the next }m_{3}\text{ measurements}\\
\varphi_{B}+\theta+\pi\text{ for the last }m_{4}\text{ measurements}%
\end{array}
\label{12}%
\end{equation}
We note that unitarity (particle conservation) requires that the third and
fourth angle are obtained by adding $\pi$ to the first and third angles). So
the time derivative of the probability of obtaining a particular sequence
$(m_{1},m_{2},m_{3},m_{4})$ with given pixels is (with the new variable
$\lambda^{\prime}=\Lambda-\lambda$, $\zeta^{\prime}=\zeta-\varphi_{A}$,
$\theta^{\prime}=\theta-\varphi_{B}$):%
\begin{equation}%
\begin{array}
[c]{l}%
\displaystyle\mathcal{P(}t\mathcal{)\sim~}%
{\displaystyle\prod\limits_{j=1}^{N}}
p_{j}(t)\int_{-\pi}^{\pi}\frac{d\lambda^{\prime}}{2\pi}~~\int_{-\pi}^{\pi
}\frac{d\Lambda}{2\pi}~~\cos\left[  (N_{\beta}-N_{\alpha})\Lambda\right]
~~\left[  \cos\Lambda+\cos\left(  \zeta^{\prime}-\lambda^{\prime}\right)
\right]  ^{m_{1}}\left[  \cos\Lambda-\cos\left(  \zeta^{\prime}-\lambda
^{\prime}\right)  \right]  ^{m_{2}}\\
\multicolumn{1}{r}{\times\left[  \cos\Lambda+\cos\left(  \theta^{\prime
}-\lambda^{\prime}\right)  \right]  ^{m_{3}}\left[  \cos\Lambda-\cos\left(
\theta^{\prime}-\lambda^{\prime}\right)  \right]  ^{m_{4}}}%
\end{array}
\label{13}%
\end{equation}

For short times, when the wave packets begin to reach the detectors, the
probabilities $p_{j}(t)$ grow linearly in time from zero, as usual in a 3
dimensional problem.\ The coincidence probability $\mathcal{P(}t\mathcal{)}$
contains a product of $N$ values of $p_{j}(t)$, so that it will initially grow
much more slowly, with only a $N$-th order non-zero time derivative.\ For
longer times, when the $p_{j}(t)$'s have grown to larger values, any
derivative of $\mathcal{P(}t\mathcal{)}$ may be non-zero.\ At the end of the
experiment, when the wave packets have entirely crossed the detectors and all
the particles are absorbed, the $p_{j}(t)$'s reach their limiting value
$\overline{p}_{j}$, and the probability is (from now on, we drop the primes,
which just introduce a redefinition of the origin of the angles) :%
\begin{equation}%
\begin{array}
[c]{l}%
\mathcal{P(}m_{1},m_{2},m_{3},m_{4})\mathcal{\sim}%
{\displaystyle\prod\limits_{j=1}^{N}}
\overline{p}_{j}\int_{-\pi}^{\pi}\frac{d\lambda}{2\pi}~~\int_{-\pi}^{\pi}%
\frac{d\Lambda}{2\pi}~~\cos\left[  (N_{\beta}-N_{\alpha})\Lambda\right]
~~\left[  \cos\Lambda+\cos\left(  \zeta-\lambda\right)  \right]  ^{m_{1}%
}\left[  \cos\Lambda-\cos\left(  \zeta-\lambda\right)  \right]  ^{m_{2}}\\
\multicolumn{1}{r}{\times\left[  \cos\Lambda+\cos\left(  \theta-\lambda
\right)  \right]  ^{m_{3}}\left[  \cos\Lambda-\cos\left(  \theta
-\lambda\right)  \right]  ^{m_{4}}}%
\end{array}
\label{13-bis}%
\end{equation}

\subsection{Counting factors and probabilities}

\label{probab}At this point, we must take counting factors into
account.\ There are:%
\begin{equation}
\frac{\mathcal{Q}!}{m_{1}!(\mathcal{Q}-m_{1})!} \label{14}%
\end{equation}
different configurations of the pixels in the first detector that lead to the
same number of detections $m_{1}$. For the two detectors in $D_{A}$, this
number becomes:%
\begin{equation}
\frac{\mathcal{Q}!}{m_{1}!(\mathcal{Q}-m_{1})!}\frac{\mathcal{Q}!}%
{m_{2}!(\mathcal{Q}-m_{2})!} \label{15}%
\end{equation}
But, if we note $m_{A}=m_{1}+m_{2}$ and use the Stirling formula, we can
approximate:%
\begin{equation}%
\begin{array}
[c]{l}%
\log(\mathcal{Q}-m_{1})!+\log(\mathcal{Q}-m_{2})!\\
\multicolumn{1}{r}{\sim\left(  \mathcal{Q}-m_{1}+\frac{1}{2}\right)  \left[
\log\mathcal{Q}+\log\left(  1-\frac{m_{1}}{\mathcal{Q}}\right)  \right]
-\left(  \mathcal{Q}-m_{1}\right)  +\left(  \mathcal{Q}-m_{2}+\frac{1}%
{2}\right)  \left[  \log\mathcal{Q}+\log\left(  1-\frac{m_{2}}{\mathcal{Q}%
}\right)  \right]  -\left(  \mathcal{Q}-m_{2}\right) }%
\end{array}
\label{16}%
\end{equation}
or, if we expand the logarithms of $(1-m_{1,2}/\mathcal{Q})$:%
\begin{equation}
\left(  2\mathcal{Q}-m_{A}+1\right)  \log\mathcal{Q}-\mathcal{Q}\frac{m_{A}%
}{\mathcal{Q}}+\frac{m_{1}^{2}+m_{2}^{2}}{\mathcal{Q}}-2\mathcal{Q}+m_{A}+...
\label{17}%
\end{equation}
the second and the fifth term cancel each other, the third can be ignored
because of (\ref{3}); an exponentiation then provides the following term in
the denominator of the counting factor:%
\begin{equation}
\frac{\left(  \mathcal{Q}!\right)  ^{2}}{\mathcal{Q}^{m_{A}}} \label{18}%
\end{equation}
The $\mathcal{Q}!$ disappear, and the number of different configurations in
region $D_{A}$ is:%
\begin{equation}
\frac{1}{m_{1}!m_{2}!}Q^{m_{A}} \label{18-b}%
\end{equation}
\qquad\qquad\

Finally, we also have to take into account the factors $\overline{p}_{j}$ in
(\ref{13-bis}).\ These factors fluctuate among all the pixel configurations we
have counted, since some pixels near the center of the modes are better
coupled to the boson field and have larger $\overline{p}_{j}$'s than those
that are on the sides.\ If we assume that the number of pixels of each
detector is much larger than $m_{1}$ and $m_{2}$, in the summation over all
possible configurations of the pixels, we can replace each $\overline{p}_{j}$
by its average $<\overline{p}>$ over the detector\footnote{If $m_{1}=1$, the
summation provides exactly $<\overline{p}>$ the average by definition.\ If
$m_{1}=2$, the second pixel can not coincide with the first, so that the
average of the product $p_{1}p_{2}$ is not exactly $<\overline{p}>^{2}$;
nevertheless, if the number of pixels $\mathcal{Q}$ is much larger than $2$,
the average is indeed $<\overline{p}>^{2}$ to a very good approximation.\ By
recurrence, as long as the number of detections $m$ remains much smaller than
the number of pixels, one can safely replace the average of the product by the
product of averages.}.\ If we assume that all detectors are identical, this
introduces a factor $<\overline{p}>^{m_{A}}$ in the counting factor.\ When we
take into account the other detection region $D_{B}$, the factor $Q^{m_{A}}$,
together with the factor $\mathcal{Q}^{m_{B}}$, can be grouped with the
prefactor $S^{N}$ in (\ref{13}) to provide $\left(  \mathcal{QS}\right)  ^{N}%
$, which contains the total detection volume to the power $N$, as
natural\footnote{The probability remains invariant if, at constant detection
area, the value of the number of pixels $\mathcal{Q}$ is increased.}; on the
other hand, the factor $<\overline{p}>^{N}$ is irrelevant, since it does not
affect the relative values.\ At the end, we recover expression (\ref{19}) for
the probability of obtaining the series of results $\mathcal{(}m_{1}%
,m_{2},m_{3},m_{4})$.

This calculation shows precisely what are the experimental parameters that are
important to preserve the interesting interference effects, and expresses them
in geometrical terms.\ The main physical idea is that the detection process
should not give any indication, even in principle, of the source from which
the particles have originated: on the detection surface, the two sources
produce indistinguishable wave functions.\ Therefore, in practice, what is
relevant is not the coherence length of the wave functions over the entire
detection regions, as the calculation of \S \ \ref{quantum} could suggest,
since in these regions the modes are defined mathematically in a half-infinite
space; what really matters is the parallelism of the wave fronts of the two
wave functions with the input surface of the detectors.\ Moreover, if
necessary, formulas such as (\ref{9}) and (\ref{10-e}) allow us to calculate
the corrections introduced by wave front mismatch, and therefore to have a
more realistic idea of the experimental requirements; for instance, if the
$\mathbf{k}_{\alpha}$ and $\mathbf{k}_{\beta}$ are not strictly parallel and
perpendicular to the surface of the photodetectors, one can write write
$\mathbf{k}_{\alpha,\beta}=\mathbf{k}_{\Delta}\pm\delta\mathbf{k}(\mathbf{r})$
and calculate the correction to (\ref{9}) to first order in $\delta\mathbf{k}%
$, etc.

\section{EPR\ argument and Bell theorem for parity}

\label{EPR}EPR variables are pairs of variables for which the result of a
measurement made by Alice can be used to predict the result of a measurement
made by Bob with certainty.\ For instance, the numbers of particles detected
by Alice and by Bob are such a pair, provided we assume that the experiment
has 100\% efficiency (no particle is missed by the detectors): Alice knows
that, if she has measured $m_{A}$ particles, Bob will detect $N-m_{A}$
particles. It is therefore possible to use the EPR argument to show that
$m_{A}$ and $m_{B}$ correspond to elements of reality that were determined
before any measurement took place.\ Moreover, this also allows us to define an
ensemble of events for which $m_{A}$ and $m_{B}$ are fixed as an ensemble that
is independent of the settings used by Alice and Bob; this independence is
essential for the derivation of the Bell inequalities within local realism
\cite{CS}.\ So we may either study situations where $m_{A}$ and $m_{B}$ are
left to fluctuate freely, or where they are fixed (as with spin condensates
\cite{PRL}).

When $N_{\alpha}=N_{\beta}$, we have seen in \S \ \ref{perfect} that another
pair of EPR\ variables is provided by the parities $\mathcal{A}=(-1)^{m_{2}}$
and $\mathcal{B}=(-1)^{m_{4}}$ of the results observed by Alice and Bob: if
they choose opposite values $\zeta=-\theta$ for their settings, perfect
correlations occur, even if Alice and Bob are at an arbitrarily large distance
from each other.\ We now study quantum violations of local realism with these variables.

\subsection{Parity and BCHSH\ inequalities}

We suppose that in the experiment of Fig. 1, Alice and Bob each use two
different angle settings, $\zeta$ and $\zeta^{\prime}$ for Alice and $\theta$
and $\theta^{\prime}$ for Bob. Within local realism (EPR\ argument), for each
realization of the experiment the observed results depend only on the local
settings.\ We can then define $\mathcal{A}$ as the parity observed by Alice if
she chooses setting $\zeta$, and $\mathcal{A}^{\prime}$ the parity if she
chooses $\zeta^{\prime}$; similarly, Bob obtains results $\mathcal{B}$ or
$\mathcal{B}^{\prime}$ depending on his choice $\theta$ or $\theta^{\prime}$;
all these results are parities equal to $\pm1.$ Then, since either
$\mathcal{B}+\mathcal{B}^{\prime}$ or $\mathcal{B}-\mathcal{B}^{\prime}$
vanishes, within local realism we have the relation:
\begin{equation}
-2\geq\mathcal{AB+AB}^{\prime}\mathcal{+A}^{\prime}\mathcal{B-A}^{\prime
}\mathcal{B}^{\prime}\leq2 \label{CHSHForm}%
\end{equation}
For an ensemble of events, the average of this quantity over many realizations
must then also have a value between $-2$ and $+2$ (BCHSH theorem).

In quantum mechanics \textquotedblleft unperformed experiments have no
results\textquotedblright\ \cite{Peres}: any given realization of the
experiment necessarily corresponds to one single whole experimental
arrangement, and it is never possible to define simultaneously all 4 numbers
in Eq. (\ref{CHSHForm}). One can nevertheless calculate the quantum average of
the product of the results for given settings, and derive the expression:
\begin{equation}
Q=\left\langle \mathcal{AB}\right\rangle +\left\langle \mathcal{AB}^{\prime
}\right\rangle +\left\langle \mathcal{A}^{\prime}\mathcal{B}\right\rangle
-\left\langle \mathcal{A}^{\prime}\mathcal{B}^{\prime}\right\rangle \label{Q}%
\end{equation}
but there is no reason to expect $Q$ to be between $-2$ and $+2$.

Since (\ref{Proba}) reduces to (\ref{19}) (with $M=N$) when $R=0$ and $T=1\,$,
we can proceed from the more general formula (\ref{Proba}).\ The calculation
of the average $\left\langle \mathcal{AB}\right\rangle $ is very similar to
that of section (iv) of the Appendix, but with a factor $(-1)^{m_{2}+m_{4}}$
included in the sum on $m_{1},\cdots,m_{4}.$ The equivalent of (\ref{app-12}),
obtained after summations over $m_{1}$ and $m_{2}$ (with constant sum $m_{A}$)
and over $m_{3}$ and $m_{4}$ (with constant sum $m_{B}$) is:%
\begin{equation}%
\begin{array}
[c]{l}%
\frac{N_{\alpha}!N_{\beta}!}{m_{A}!m_{B}!}\frac{2^{N-2M}}{\left(  N-M\right)
!}T^{M}R^{N-M}~\int_{-\pi}^{+\pi}\frac{d\Lambda}{2\pi}\cos\left[  \left(
N_{\alpha}-N_{\beta}\right)  \Lambda\right]  \left[  \cos\Lambda\right]
^{N-M}\\
\multicolumn{1}{r}{\times\int_{-\pi}^{\pi}\frac{d\lambda}{2\pi}\left[
2\cos\left(  \zeta+\lambda\right)  \right]  ^{m_{A}}\left[  2\cos\left(
\theta-\lambda\right)  \right]  ^{m_{B}}}%
\end{array}
\label{P-1}%
\end{equation}
Formula (\ref{app-8}) of the Appendix can then be used, with $M$ replaced by
$N-M$.\ Therefore we see that the average $\left\langle \mathcal{AB}%
\right\rangle $ of the product vanishes, unless the two following conditions
are met:%
\begin{equation}
\left\{
\begin{array}
[c]{l}%
M\text{ is even}\\
M\leq2N_{\alpha}\text{ and }M\leq2N_{\beta}%
\end{array}
\right.  \label{P-2}%
\end{equation}
It these two conditions are met, the first line of (\ref{P-1}) becomes:%
\begin{equation}
\frac{N_{\alpha}!N_{\beta}!}{m_{A}!m_{B}!}~2^{-M}\frac{T^{M}R^{N-M}}{\left(
N_{\alpha}-\frac{M}{2}\right)  !\left(  N_{\beta}-\frac{M}{2}\right)  !}
\label{P-3}%
\end{equation}
while the second line provides, with the help of formula (\ref{app-3}) of the
Appendix:%
\begin{equation}
2^{M}\left[  \cos\left(  \frac{\zeta+\theta}{2}\right)  \right]  ^{M}%
\int_{-\pi}^{\pi}\frac{d\lambda}{2\pi}\left[  \cos\left(  \lambda+\frac
{\zeta-\theta}{2}\right)  \right]  ^{M}=~M!\left[  \left(  \frac{M}{2}\right)
!\right]  ^{-2}\left[  \cos\left(  \frac{\zeta+\theta}{2}\right)  \right]
^{M} \label{P-4}%
\end{equation}
Finally, the sum over $m_{A}$ and $m_{B}~$(with constant sum $M$) gives the
result:%
\begin{equation}
\left\langle \mathcal{AB}\right\rangle =\frac{N_{\alpha}!N_{\beta}!}{\left(
N_{\alpha}-\frac{M}{2}\right)  !\left(  N_{\beta}-\frac{M}{2}\right)  !\left[
\left(  \frac{M}{2}\right)  !\right]  ^{2}}T^{M}R^{N-M}\left[  \cos\left(
\frac{\zeta+\theta}{2}\right)  \right]  ^{M} \label{P-5}%
\end{equation}

If $M$ is left to fluctuate, a summation of this expression over $M$ should be
done. But another point of view is to decide to count only the events where
$M$ is fixed\footnote{With the experimental setup of fig.\ \ref{fig2}, one can
include the results of measurements given by the detectors in channels 5 and 6
in the preparation procedure; only the events in which $m_{5}+m_{6}=N-M$ are
retained in the sample considered.\ Since this procedure remains independent
of the settings $\theta$ and $\zeta$, this does not open a \textquotedblleft
sample bias loophole\textquotedblright.}.\ Then this average should be
compared with the probability that $M$ particles will be detected, given by
formula (\ref{app-14}) in the Appendix; dividing (\ref{P-5}) by (\ref{app-14})
now provides:%
\begin{equation}
\left\langle \mathcal{AB}\right\rangle =\frac{N_{\alpha}!N_{\beta}%
!M!(N-M)!}{N!\left(  N_{\alpha}-\frac{M}{2}\right)  !\left(  N_{\beta}%
-\frac{M}{2}\right)  !\left[  \left(  \frac{M}{2}\right)  !\right]  ^{2}%
}\left[  \cos\left(  \frac{\zeta+\theta}{2}\right)  \right]  ^{M}
\label{P-5-bis}%
\end{equation}

In the case $M=N,$ the second condition (\ref{P-2}) requires than $N_{\alpha
}=N_{\beta}=M/2$, in which case we get:
\begin{equation}
\left\langle \mathcal{AB}\right\rangle =\cos\left(  \frac{\zeta+\theta}%
{2}\right)  ^{N}\text{ \ \ for~}~\text{~}~M=N~\text{~}~\text{~}~ \label{P-6}%
\end{equation}
One can put this into $Q$ of Eq. (\ref{Q}): Alice's measurement angle is taken
for convenience as $\phi_{a}=2\zeta$ and Bob's as $\phi_{b}=-2\theta.$ Then
defining $E(\phi_{a}-\phi_{b})=\cos^{N}\left(  \phi_{a}-\phi_{b}\right)  $ and
setting $\phi_{a}-\phi_{b}=\phi_{b}-\phi_{a^{\prime}}=\phi_{b^{\prime}}%
-\phi_{a}=\omega$ and $\phi_{b^{\prime}}-\phi_{a^{\prime}}=3\omega$ we can
maximize $Q=3E(\omega)-E(3\omega)$ to find the greatest violation of the
inequality for each $N$. For $N=2$ we find $Q_{\max}=2.41$ in agreement with
Ref. \cite{YS-1}; for $N=4,$ $Q_{\max}=2.36;$ and for $N\rightarrow\infty,$
$Q_{\max}\rightarrow2.33$. These values are obtained for a value of the angles
corresponding to $\omega=\sqrt{\ln3/N}$, which decreases relatively slowly
with $N$. The conclusion is that the system continues to violate local realism
for \emph{arbitrarily large condensates}. As already noted in \S \ \ref{par},
this is a direct consequence of the effects of the quantum angle $\Lambda$,
since no such violation could occur if this angle was zero.

Suppose now we measure $M=N-1$ particles with $N_{\beta}=M/2,$ $N_{\alpha
}=M/2+1.$ Then the coefficient of the cosine in Eq. (\ref{P-5-bis}) is
$(M/2+1)/(M+1)$, which is 2/3 at $M=2$ and smaller for larger $M,$ so that
this case never violates the BCHSH inequality since $2/3\times2.41=1.61<2$. If
(\ref{P-5}) had been used instead of (\ref{P-5-bis}), we would be even even
further from any violation, since the first average value is smaller than the
second.\ The conclusion is that \emph{even one single missed particle ruins
the quantum violation}.

\subsection{Three Fock states and three interferometers; GHZ\ contradictions}

\label{Three Fock}
\begin{figure}[h]
\centering\includegraphics[width=14pc]{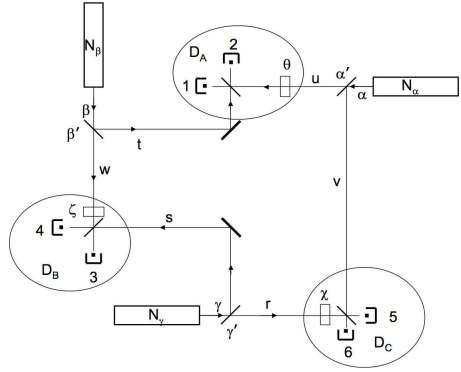}\hspace{2pc}%
\caption{Interferometer with three Fock-state condensate sources and three
detectors. The particles from each source can reach two detectors. Each
detector has two subdetectors, which will register a $+1$ for the odd-numbered
subdetector and $-1$ for the even-numbered. We average the quantity
$\mathcal{ABC}$ where $\mathcal{A}=\pm1$ for Alice's detector, and
$\mathcal{B}=\pm1$ for Bob's, and $\mathcal{C}=\pm1$ for Carole's.}%
\label{fig3}%
\end{figure}

With a triple-Fock state source (TFS) as shown in
Fig.\ref{fig3} we can demonstrate GHZ contradictions \cite{GHZ-1, GHZ-2}. Such
a contradiction occurs when local realism predicts a quantity to be, say, $+1$
while quantum mechanics predicts the opposite, $-1.$ Previous such
contradictions were carried out with states known variously as GHZ states,
NOON states, or maximally entangled states. These wave functions are of the
form $u\left\vert +++\cdots\right\rangle +v\left\vert ---\cdots\right\rangle $
with particular values of the phases $u$ and $v.$ The original GHZ
calculations \cite{GHZ-1} were done with three- and four-body NOON\ states,
and this was generalized to $N$ particles by Mermin \cite{Mermin}. Yurke and
Stoler \cite{YS-2} showed how an interferometer with three one-particle
sources also could give a GHZ contradiction. We will replace their sources
with Bose condensates to show how new $N$-body contradictions can be developed.

The initial TFS is:
\begin{equation}
\left\vert \Phi\right\rangle ~=\left\vert N_{\alpha},N_{\beta},N_{\gamma
}\right\rangle =\frac{1}{\sqrt{N_{\alpha}!N_{\beta}!N_{\gamma}!}}a_{\alpha
}^{\dagger N_{\alpha}}a_{\beta}^{\dagger N_{\beta}}a_{\gamma}^{\dagger
N_{\gamma}}\left\vert 0\right\rangle \label{TFS}%
\end{equation}
As in \S \ \ref{quantum}, the output modes (destruction operators)
$a_{1}\cdots a_{6}$ can be written in terms of the modes at the sources
$a_{\alpha},a_{\beta}$ and $a_{\gamma}$ with three phase shifts of $\zeta,$
$\chi,$ or $\theta$. We find:
\begin{align}
a_{1}  &  =\frac{1}{2}\left[  e^{i\zeta}a_{\alpha}-ia_{\beta}\right]
,~\text{~}~\text{~}~\text{~}~a_{2}=\frac{1}{2}\left[  ie^{i\zeta}a_{\alpha
}-a_{\beta}\right]  ,\nonumber\\
a_{3}  &  =\frac{1}{2}\left[  e^{i\theta}a_{\beta}-ia_{\gamma}\right]
,~\text{~}~\text{~}~\text{~}~a_{4}=\frac{1}{2}\left[  ie^{i\theta}a_{\beta
}-a_{\gamma}\right]  ,\nonumber\\
a_{5}  &  =\frac{1}{2}\left[  -a_{\alpha}+e^{i\chi}a_{\gamma}\right]
,\text{~~}~\text{~}~\text{~}a_{6}=\frac{1}{2}\left[  ia_{\alpha}+ie^{i\chi
}a_{\beta}\right]  .
\end{align}
We write generally $a_{i}=v_{i\alpha}a_{\alpha}+v_{i\beta}a_{\beta}%
+v_{i\gamma}a_{\gamma}$. We consider only the case where every particle in the
source is detected, so the probability that we find $m_{i}$ particles in
detector $i=1\cdots6,$ is:
\begin{equation}
\mathcal{P(}m_{1},\cdots,m_{6})\sim\frac{1}{m_{1}!\cdots m_{6}!}\left\vert
\left\langle 0\right\vert a_{1}^{m_{1}}\cdots a_{6}^{m_{6}}\left\vert
N_{\alpha},N_{\beta},N_{\gamma}\right\rangle \right\vert ^{2}%
\end{equation}
(this relation is actually an equality, but we write it only as a
proportionality relation since we will change the normalization below).\ We
can develop the matrix element just as we did in \S \ \ref{PVR}:%
\begin{equation}%
\begin{array}
[c]{l}%
\displaystyle\left\langle 0\right\vert \prod_{i=1}^{6}\left(  v_{i\alpha
}a_{\alpha}+v_{i\beta}a_{\beta}+v_{i\gamma}a_{\gamma}\right)  ^{m_{i}%
}a_{\alpha}^{\dagger N_{\alpha}}a_{\beta}^{\dagger N_{\beta}}a_{\gamma
}^{\dagger N_{\gamma}}\left\vert 0\right\rangle =N_{\alpha}!N_{\beta
}!N_{\gamma}!~\times\\
\multicolumn{1}{r}{\displaystyle\times\sum_{p_{1=0}}^{m_{1}}...\sum_{p_{6=0}%
}^{m_{6}}\left[  \prod_{i=1}^{6}\left(  \frac{m_{i}!}{p_{i\alpha}!p_{i\beta
}!p_{i\gamma}!}u_{i\alpha}^{p_{i\alpha}}u_{i\beta}^{p_{i\beta}}u_{i\gamma
}^{p_{i\gamma}}\right)  ~\delta_{p_{1\alpha}+\cdots+p_{6\alpha},N_{\alpha}%
}~\delta_{p_{1\beta}+\cdots+p_{6\beta},N_{\beta}}~\delta_{p_{1\gamma}%
+\cdots+p_{6\gamma},N_{\gamma}}\right] }%
\end{array}
\label{number}%
\end{equation}
where the sums are over all $p_{i\alpha}$, $p_{i\beta}$ and $p_{i\gamma}$ such
that $p_{i\alpha}+p_{i\beta}+p_{i\gamma}=m_{i}$. We now replace the $\delta
$-functions by integrals:
\begin{equation}
\delta_{p_{1\alpha}+\cdots+p_{6\alpha},N_{\alpha}}=\int_{-\pi}^{\pi}%
\frac{d\lambda_{\alpha}}{(2\pi)^{3}}e^{i(p_{1\alpha}+\cdots+p_{6\alpha
}-N_{\alpha})\lambda_{\alpha}}%
\end{equation}
with similar integrals over $\lambda_{\beta}$ and $\lambda_{\gamma}.$ In the
sum above then, we have every $v_{i\alpha}^{p_{i\alpha}}$ replaced by $\left(
v_{i\alpha}e^{i\lambda_{\alpha}}\right)  ^{p_{i\alpha}}$ etc. so that we can
redo the sums over the $p_{i\alpha},$ etc. to find the probability for the
$m_{i}$ arrangement under the condition that all the source particles are
detected:
\begin{equation}
\mathcal{P(}m_{1},\cdots,m_{6})\sim\frac{1}{m_{1}!\cdots m_{6}!}\int
d\tau^{\prime}\int d\tau e^{-i[N_{\alpha}(\lambda_{\alpha}-\lambda_{\alpha
}^{\prime})+N_{\beta}(\lambda_{\beta}-\lambda_{\beta}^{\prime})+N_{\gamma
}(\lambda_{\gamma}-\lambda_{\gamma}^{\prime})]}\prod_{i=1}^{6}\left(
\Omega_{i}^{\prime\ast}\Omega_{i}\right)  ^{m_{i}} \label{Proba-1-6}%
\end{equation}
where $\Omega_{i}$ $=v_{i\alpha}e^{i\lambda_{\alpha}}+v_{i\beta}%
e^{i\lambda_{\beta}}+v_{i\gamma}e^{i\lambda_{\gamma}}$ and $\Omega_{i}%
^{\prime}$ has the same expression with primed $\lambda$'s; $d\tau$ represents
the integrals over $\lambda_{\alpha},$ $\lambda_{\beta},$ and $\lambda
_{\gamma},$ and $d\tau^{\prime}$ over the $\lambda_{\alpha}^{\prime},$
$\lambda_{\beta}^{\prime},$ and $\lambda_{\gamma}^{\prime}.$

In a ideal experiment with 100\% detection efficiency, the numbers of
particles detected in each region are EPR\ variables, since the value of two
of these variables determines the value of the third with certainty; these
perfect correlations are independent of the settings (phase shifts of $\zeta,$
$\chi,$ or $\theta$), so that choosing the number of detections in each region
defines a class of events that is independent of the settings. Here, assuming
that each source emits $N/3$ particles (otherwise, we find zero average
values, see below):%
\[
N_{\alpha}=N_{\beta}=N_{\gamma}=N/3
\]
we will also assume that each detector registers exactly $N/3$
particles\footnote{We have also performed more general calculations---not
given here---in which this restriction does not apply, but then we have found
no GHZ contradictions.}. We can put in this restriction, when we sum on
$m_{1}\cdots m_{6}$ to get averages, by including three $\delta$-functions of
the form:
\begin{equation}
\delta_{m_{1}+m_{2},N/3}=\int_{-\pi}^{\pi}\frac{d\rho_{A}}{2\pi}~e^{i\rho
_{A}(m_{1}+m_{2}-N/3)}%
\end{equation}
with similar ones specifying $m_{3}+m_{4}=N/3$ and $m_{5}+m_{6}=N/3$. The
$m_{i}$ sums are then done independently of one another giving a normalization
sum of:
\begin{align}
\mathcal{N}  &  =\int d\tau_{\rho}\int d\tau^{\prime}\int d\tau
e^{-i[N_{\alpha}(\lambda_{\alpha}-\lambda_{\alpha}^{\prime})+N_{\beta}%
(\lambda_{\beta}-\lambda_{\beta}^{\prime})+N_{\gamma}(\lambda_{\gamma}%
-\lambda_{\gamma}^{\prime})]}\nonumber\\
&  \times e^{-iN/3[\rho_{A}+\rho_{B}+\rho_{C}]}\exp\left[  \sum_{i=1}%
^{6}\left(  \Omega_{i}^{\prime\ast}\Omega_{i}e^{i\rho_{i}}\right)  \right]
\end{align}
where $\rho_{1}=\rho_{2}=\rho_{A},$ $\rho_{3}=\rho_{4}=\rho_{B},$ and
$\rho_{5}=\rho_{6}=\rho_{C}$ and $\int d\tau_{\rho}$ represents the new
three-fold integration. The sum in the exponential is easily done:
\begin{align}
\sum_{i=1}^{6}\left(  \Omega_{i}^{\prime\ast}\Omega_{i}e^{i\rho_{i}}\right)
&  =\frac{1}{2}\left[  e^{-i(\lambda_{\alpha}-\lambda_{\alpha}^{\prime}%
)}\left(  e^{i\rho_{A}}+e^{i\rho_{C}}\right)  \right. \nonumber\\
&  \left.  +e^{-i(\lambda_{\beta}-\lambda_{\beta}^{\prime})}\left(
e^{i\rho_{B}}+e^{i\rho A}\right)  +e^{i(\lambda_{\gamma}-\lambda_{\gamma
}^{\prime})]}\left(  e^{i\rho_{C}}+e^{i\rho_{B}}\right)  \right]
\end{align}
We expand the exponential of this quantity in series in $e^{-i(\lambda
_{\alpha}-\lambda_{\alpha}^{\prime})},$ $e^{-i(\lambda_{\beta}-\lambda_{\beta
}^{\prime})},$ and $e^{i(\lambda_{\gamma}-\lambda_{\gamma}^{\prime})}$ and do
the integrals. When each source emits exactly $N/3$ particles, we obtain:
\begin{equation}
\mathcal{N}=\frac{1}{2^{N}}\sum_{l=0}^{\infty}\left(  \frac{1}{l!(\frac{N}%
{3}-l)!}\right)  ^{3}%
\end{equation}
Similarly we average the quantities $\mathcal{A},$ $\mathcal{B},$ and
$\mathcal{C}$ each $\pm1$ measured by Alice, Bob, and Carole according to:
\begin{equation}
\left\langle \mathcal{ABC}\right\rangle =\sum_{m_{1}\cdots m_{6}}{}^{\prime
}(-1)^{m_{2}+m_{4}+m_{6}}\mathcal{P(}m_{1},\cdots,m_{6})
\end{equation}
where the prime on the sum means we again restrict the sums to the case of
$N/3$ particles reaching each detector. With this requirement the average
vanishes unless each source emits exactly $N/3$ particles, which is why above
we considered just that case. We have then:
\begin{equation}
\left\langle \mathcal{ABC}\right\rangle =\frac{\sum_{q}\left(  \frac
{N/3!}{(N/3-q)!q!}~\right)  ^{3}e^{i(\zeta+\theta+\chi)(N/3-2q)}}{\sum
_{q}\left(  \frac{N/3!}{(N/3-q)!q!}~\right)  ^{3}} \label{GenGHZResult}%
\end{equation}
Note that if $\zeta+\theta+\chi=0$ we find $\left\langle \mathcal{ABC}%
\right\rangle =1$: perfect correlations exist between the results since their
product is fixed.\ Thus, if we know the parity of the results of two of the
experimenters, we immediately know that of the third, even if that person is
very far away. Thus an EPR argument applies to these variables.

For the case $N=3,$ we find $\left\langle \mathcal{ABC}\right\rangle
=\cos(\zeta+\theta+\chi),$ which is the same form of the original GHZ case
found from a three-body NOON\ state. This result agrees with the
interferometer result of Ref. \cite{YS-2} as expected. Local realism predicts
that, for each realization of the experiment, the product of the results is
given by a product $A(\zeta)B(\theta)C(\chi)$. To get agreement with quantum
mechanics in situations of perfect correlations we must have:
\begin{align}
A(\pi/2)~B(\pi/2)~C(0)  &  =-1\nonumber\\
A(\pi/2)~B(0)~C(\pi/2)  &  =-1\nonumber\\
A(0)~B(\pi/2)~C(\pi/2)  &  =-1
\end{align}
But then we obtain by product $A(0)B(0)C(0)=-1,$ while quantum mechanics gives
$+1$, in complete contradiction. In our case we get new contradictions for
larger $N$; consider for instance $N=9,$ in which case:
\begin{equation}
\left\langle \mathcal{ABC}\right\rangle =\frac{1}{28}\left[  27\cos
(\zeta+\theta+\chi)+\cos3(\zeta+\theta+\chi)\right]
\end{equation}
The above argument goes through exactly in the same way. More generally, any
time $N/3$ is odd we get a similar result for arbitrary $N.$

Thus the TFS provides new GHZ-type contradictions for $N$ particles
\emph{without} having to prepare NOON states.

\subsection{Hardy impossibilities}

\begin{figure}[h]
\centering \includegraphics[width=2.5in]{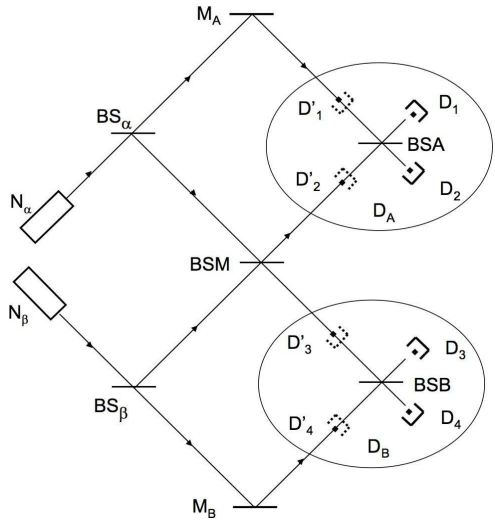}\caption{An interferometer
with particle sources $\alpha$ and $\beta$, with beam splitters designated by
BS and mirrors by M. In both detection regions, the detectors at D$_{i}$ may
be replaced by the D$_{i}^{\prime}$, placed before the beam splitters$.$ }%
\label{fig4}%
\end{figure}

Hardy impossibilities are treated by use of the interferometer shown in
Fig.~\ref{fig4}, based on the one discussed in Ref.~\cite{H-1} for $N=2$.
The heart of the system is the beam splitter at the center; due to
Bose interference it has the property that, if an equal number of particles
approaches each side, then an even number must emerge from each side. The
detection beam splitters BSA and BSB are each set to have a transmission
probability of 1/3, and the path differences are such that, by destructive
interference, no particle reaches $D_{2}$ if only source $N_{\alpha}$ is used;
similarly, no particle reaches $D_{3}$ if $N_{\beta}$ alone is used. Alice can
use either the detectors $D_{1,2}$ after her beam splitter, or $D_{1,2}%
^{\prime}$ before; Bob can choose either D$_{3,4}$, or D$_{3,4}^{\prime}$.
This gives $4$ arrangements of experiments: $DD$, $DD^{\prime}$, $D^{\prime}%
D$, or $D^{\prime}D^{\prime}$, with probability amplitudes $C_{XY}(m_{1}%
,m_{2};m_{3},m_{4})$, where $XY$ is any of these $4$ arrangements and the $m$
values are the numbers of particles detected at each counter.

We find the destruction operators for the detector modes as we have done in
previous sections. For the primed detectors we find:%
\begin{equation}%
\begin{array}
[c]{l}%
a_{D_{1}^{\prime}}=\frac{i}{\sqrt{2}}a_{\alpha},~\text{~}~\text{~}%
~\text{~}~\text{~}~\text{~}~\text{~}~\text{~}a_{D_{2}^{\prime}}=\frac{1}%
{2}\left(  -a_{\alpha}+ia_{\beta}\right) \\
a_{D_{3}^{\prime}}=\frac{1}{2}\left(  ia_{\alpha}-a_{\beta}\right)
,~\text{~}~a_{D_{4}^{\prime}}=\frac{i}{\sqrt{2}}a_{\beta}%
\end{array}
\label{eqn}%
\end{equation}
and for the unprimed detectors:%
\begin{equation}%
\begin{array}
[c]{l}%
a_{D_{1}}=-\frac{\sqrt{3}}{2}a_{\alpha}+\frac{i}{2\sqrt{3}}a_{\beta}%
,~\text{~}~\text{~}~\text{~}~a_{D_{2}}=-\frac{1}{\sqrt{6}}a_{\beta}\\
a_{D_{3}}=-\frac{1}{\sqrt{6}}a_{\alpha},~\text{~}~\text{~}~\text{~}%
~\text{~}~\text{~}~\text{~}~\text{~}~\text{~}~\text{~}~\text{~}~a_{D_{4}%
}=\frac{i}{2\sqrt{3}}a_{\alpha}-\frac{\sqrt{3}}{2}a_{\beta}%
\end{array}
\label{equat}%
\end{equation}
In general we write these results as:
\begin{equation}
a_{i}=v_{i\alpha}a_{\alpha}+v_{i\beta}a_{\beta}%
\end{equation}
Note that, because of the $1/3$ transmission probability at BSA and BSB, Bose
interference causes $a_{\alpha}$ and $a_{\beta}$ to drop out of the second and
third of Eqs (\ref{equat}), respectively.

The amplitude is given by:
\begin{equation}
C_{\mathrm{XY}}(m_{1},m_{2};m_{3},m_{4})\sim\left\langle 0\right\vert
\prod_{i=1}^{6}\left(  v_{i\alpha}a_{\alpha}+v_{i\beta}a_{\beta}\right)
^{m_{i}}a_{\alpha}^{\dagger N_{\alpha}}a_{\beta}^{\dagger N_{\beta}}\left\vert
0\right\rangle
\end{equation}
As we have done in previous sections, we expand the binomials, evaluate the
operator matrix element, replace the resulting $\delta$-functions by integrals
and resum the series to find:
\begin{equation}
C_{\mathrm{XY}}(m_{1},m_{2};m_{3},m_{4})\sim\int_{-\pi}^{\pi}\frac
{d\lambda_{\alpha}}{2\pi}\int_{-\pi}^{\pi}\frac{d\lambda_{\beta}}{2\pi
}e^{^{-iN_{\alpha}\lambda_{\alpha}}}e^{^{-iN_{\beta}\lambda_{\beta}}}%
\prod_{i=1}^{4}\left(  v_{i\alpha}e^{i\lambda_{\alpha}}+v_{i\beta}%
e^{i\lambda_{\beta}}\right)  ^{m_{i}}%
\end{equation}

In all the following we assume that each source emits $N/2$ particles, where
$N/2$ is \emph{odd}, and that detector A and detector B each receive exactly
$N/2$ particles; this is possible since, as above, the number of particles
detected in each region can define a sample of realizations that is
independent of the settings (we have to make this assumption since it turns
out that the argument works only in this case). First consider both Alice and
Bob using primed detectors. The amplitude for receiving $N/2$ particles in
each of D$_{2}^{\prime}$ and D$_{3}^{\prime}$ is:
\begin{align}
C_{\mathrm{D}^{\prime}\mathrm{D}^{\prime}}(0,\frac{N}{2};0,\frac{N}{2})  &
\sim\int_{-\pi}^{\pi}\frac{d\lambda_{\alpha}}{2\pi}\int_{-\pi}^{\pi}%
\frac{d\lambda_{\beta}}{2\pi}e^{^{-i\frac{N}{2}(\lambda_{\alpha}%
+\lambda_{\beta})}}\left(  -e^{i\lambda_{\alpha}}+ie^{i\lambda_{\beta}%
}\right)  ^{N/2}\left(  ie^{i\lambda_{\alpha}}-e^{i\lambda_{\beta}}\right)
^{N/2}\nonumber\\
&  \sim\int_{-\pi}^{\pi}\frac{d\lambda_{\alpha}}{2\pi}\int_{-\pi}^{\pi}%
\frac{d\lambda_{\beta}}{2\pi}e^{^{-i\frac{N}{2}(\lambda_{\alpha}%
+\lambda_{\beta})}}\left(  e^{i2\lambda_{\alpha}}+e^{i2\lambda_{\beta}%
}\right)  ^{N/2}=0
\end{align}
This quantity must vanish because $N/2$ is odd. This situation is an example
of the beam splitter rule mentioned above. The result is that D$_{2}^{\prime}$
and D$_{3}^{\prime}$ cannot collect all the particles if $N/2$ are detected on
each side.

Consider next the case where one experimenter uses a primed set of detectors
and the other the unprimed:
\begin{align}
C_{\mathrm{DD}^{\prime}}(0,\frac{N}{2};m_{3}^{\prime},m_{4}^{\prime})  &
\sim\int_{-\pi}^{\pi}\frac{d\lambda_{\alpha}}{2\pi}\int_{-\pi}^{\pi}%
\frac{d\lambda_{\beta}}{2\pi}e^{^{-i\frac{N}{2}(\lambda_{\alpha}%
+\lambda_{\beta})}}\left(  e^{i\lambda_{\beta}}\right)  ^{N/2}\left(
ie^{i\lambda_{\alpha}}-e^{i\lambda_{\beta}}\right)  ^{m_{3}^{\prime}}\left(
e^{i\lambda_{\beta}}\right)  ^{m_{4}^{\prime}}\nonumber\\
&  \sim\delta_{m_{4}^{\prime},0} \label{DD'A}%
\end{align}
The factor $\left(  e^{i\lambda_{\beta}}\right)  ^{N/2}$ combines with the
first exponential so that $m_{4}^{\prime}$ must vanish. This quantity vanishes
because of the destructive interference effects at BSA and BSB caused by the
1/3 transmission probability of the beam splitters; but $C_{\mathrm{DD}%
^{\prime}}(0,\frac{N}{2};\frac{N}{2},0)\neq0$. Thus, if Alice observes $N/2$
particles at D$_{2}$, when Bob uses the primed detectors he observes with
certainty $N/2$ particles at D$_{3}^{\prime}$; similarly, if Bob has seen
$N/2$ particles in D$_{3}$, in the D$^{\prime}$D configuration Alice must see
$N/2$ in D$_{2}^{\prime}$ .

We now consider events where both experimenters do unprimed experiments and
each of them finds $N/2$ particles in D$_{2}$ and D$_{3}$; the corresponding
probability is:%
\begin{equation}
C_{\mathrm{DD}}(0,\frac{N}{2};\frac{N}{2},0)\sim\int_{-\pi}^{\pi}%
\frac{d\lambda_{\alpha}}{2\pi}\int_{-\pi}^{\pi}\frac{d\lambda_{\beta}}{2\pi
}e^{^{-i\frac{N}{2}(\lambda_{\alpha}+\lambda_{\beta})}}\left(  e^{i\lambda
_{\beta}}\right)  ^{N/2}\left(  e^{i\lambda_{\alpha}}\right)  ^{N/2}\neq0,
\end{equation}
(for $N=6$, the normalized value is $1/216$), which means that events exist
where $N/2$ particles are detected at both detectors D$_{2}$ and D$_{3}$.
However, in any of these events, if Bob had at the last instant changed to the
primed detectors, he would surely have obtained $N/2$ particles in
D$_{3}^{\prime}$, because of the certainty mentioned above (while Alice still
has $N/2$ particles in D$_{2}$). Similarly, if it is Alice who chooses primed
detectors at the last moment, she always obtains $N/2$ particles in
D$_{2}^{\prime}$ (while Bob continues to have $N/2$ particles in D$_{3}$).
Now, had both changed their minds after the emission and chosen the primed
arrangement, local realism implies that they would have found $N/2$ particles
each in D$_{2}^{\prime}$ and D$_{3}^{\prime}$: such events must exist.\ But
the corresponding quantum probability is zero, in complete contradiction. The
result is the Hardy impossibility of Ref. \cite{H-1} generalized to $N$ particles.

\section{Conclusion}

Fock-state condensates appear as remarkably versatile, able to create
violations that usually require elaborate entangled wave functions, and
produce new $N$-body violations. Compared to GHZ states or other elaborate
quantum states, they have the advantage of being accessible through the
phenomenon of Bose-Einstein condensation, with no limitation in principle
concerning the number of particles involved.\ By contrast, the production of
GHZ states requires elaborate measurement procedures, so that it seems
difficult to produce them with more than a few particles (to our knowledge,
the present world record is 5, see \cite{GHZ-5}); moreover, they are much more
sensitive to decoherence, which destroys their quantum coherence properties
\cite{DBB}.

From an experimental point of view, the major requirement is that all
particles present in the initial double Fock state should be detected, which
will of course put a practical limit on the number of particles
involved.\ Using Bose condensed gases of metastable He atoms seems to be an
attractive possibility, since the detection of individual atoms is possible
with micro-channel plates \cite{Saubamea, Robert}.\ With alkali atoms, one
could also measure the position of the particles at the outputs of
interferometers by laser fluorescence, obtaining a non-destructive quantum
measurement of $m_{1}$, ..$m_{4}$. The realization of interferometers seems
also possible, since interferometry with Bose-Einstein condensates has already
been performed \cite{BS-1} with the help of Bragg scattering optical beam
splitters \cite{BS-2, BS-3}.\ Another possibility may be to use cavity quantum
electrodynamics and quantum non-demolition photon counting methods
\cite{Guerlin} to prepare multiple Fock states.\ Experiments therefore do not
seem to be out of reach.

Laboratoire Kastler Brossel is \textquotedblleft UMR 8552 du CNRS, de l'ENS,
et de l'Universit\'{e} Pierre et Marie Curie\textquotedblright.

\bigskip

\begin{center}
APPENDIX I
\end{center}

In this appendix, we give some formulas that are useful for the calculations
of this article, in particular to check the normalization of (\ref{19}) and
(\ref{Proba}).

(i) Wallis integral.\ We consider the integral:%
\begin{equation}
K=\int_{-\pi}^{+\pi}\frac{d\lambda}{2\pi}~\left[  \cos\lambda\right]  ^{N}
\label{app-1}%
\end{equation}
which, if the limits are changed to $0$ and $\pi/2$, becomes a Wallis integral
(divided by $2\pi$ if one takes the usual definition of these integrals).
Expanding the integrand provides:%
\begin{equation}
2^{-N}\left[  e^{i\lambda}+e^{-i\lambda}\right]  ^{N}=2^{-N}\sum_{q=0}%
^{N}\frac{N!}{q!(N-q)!}~e^{i\left(  N-2q\right)  \lambda} \label{app-2}%
\end{equation}
The only exponentials that survive the $\lambda$ integration are those with
vanishing exponent ($N-2q=0$). Therefore:%
\begin{equation}%
\begin{array}
[c]{l}%
\text{if }N\text{ is odd, }K=0\\
\text{if }N\text{ is even, }K=2^{-N}\frac{N!}{\left[  (N/2)!\right]  ^{2}}%
\end{array}
\label{app-3}%
\end{equation}

(ii) Normalization integral.\ We define:%
\begin{equation}
J=\int_{-\pi}^{+\pi}\frac{d\Lambda}{2\pi}\cos\left[  \left(  N_{\alpha
}-N_{\beta}\right)  \Lambda\right]  \left[  \cos\Lambda\right]  ^{M}%
=\operatorname{Re}\left\{  \int_{-\pi}^{+\pi}\frac{d\Lambda}{2\pi}e^{i\left(
N_{\alpha}-N_{\beta}\right)  \Lambda}\left[  \cos\Lambda\right]  ^{M}\right\}
\label{app-4}%
\end{equation}
and expand:%
\begin{equation}
\left[  \cos\Lambda\right]  ^{M}=\left[  \frac{e^{i\Lambda}+e^{-i\Lambda}}%
{2}\right]  ^{M}=2^{-M}\sum_{q=0}^{M}\frac{M!}{q!(M-q)!}~e^{i\left(
M-2q\right)  \lambda} \label{app-5}%
\end{equation}
Only a term with $M-2q=N_{\alpha}-N_{\beta}$ can survive the integration, so
that:%
\begin{equation}
q=\frac{N_{\beta}-N_{\alpha}-M}{2} \label{app-6}%
\end{equation}

Therefore, $J$ is non-zero only if:%
\begin{equation}
M\text{ has the same parity as }N_{\alpha}-N_{\beta}\text{ \ ; \ }-M\leq
N_{\alpha}-N_{\beta}\leq+M \label{app-7}%
\end{equation}
and then:%
\begin{equation}
J=2^{-M}\frac{M!}{\left(  \frac{M+N_{\alpha}-N_{\beta}}{2}\right)  !\left(
\frac{M-N_{\alpha}+N_{\beta}}{2}\right)  !} \label{app-8}%
\end{equation}

(iii) Normalization of (\ref{19}).\ We now consider the probabilities
$\mathcal{P(}m_{1},m_{2},m_{3},m_{4})$ given by (\ref{19}) and calculate their
sum over $m_{1},m_{2},m_{3},m_{4}$, when these variables have a constant sum
$N$.\ We do these sums in three steps: a summation over $m_{1}$ and $m_{2}$
(with constant sum $m_{A}$), a summation over $m_{3}$ and $m_{4}$ (with
constant sum $m_{A}$), and a summation over $m_{A}$ and $m_{B}$ (with constant
sum $M$). The first two sums reconstruct powers of a binomial, the $\lambda$
integral disappears, and we obtain:%
\begin{equation}
N_{\alpha}!N_{\beta}!~2^{-N}\int_{-\pi}^{+\pi}\frac{d\Lambda}{2\pi}\cos\left[
\left(  N_{\alpha}-N_{\beta}\right)  \Lambda\right]  \times\frac{1}%
{m_{A}!m_{B}!}\left[  2\cos\Lambda\right]  ^{N} \label{app-9}%
\end{equation}
which, with (\ref{app-8}) for $M=N=N_{\alpha}+N_{\beta}$, gives:%
\begin{equation}
2^{-N}\text{~}\times\frac{M!}{m_{A}!m_{B}!} \label{app-10}%
\end{equation}
Then the summation over $m_{A}$ and $m_{B}$ gives:%
\begin{equation}
2^{-N}\left(  1+1\right)  ^{N}=1 \label{app-11}%
\end{equation}
as expected.

(iv) Normalization of (\ref{Proba}).\ We now consider the probabilities
$\mathcal{P(}m_{1},m_{2},m_{3},m_{4})$ given by (\ref{Proba}) and calculate
their sum over\ any $m_{1},m_{2},m_{3},m_{4}$.\ We do this by the same three
summation as above (with $m_{A}+m_{B}=M$, instead of $N$), plus a summation
over $M$ ranging from $0$ to $N$. The first two summations give:%
\begin{equation}
\frac{N_{\alpha}!N_{\beta}!}{m_{A}!m_{B}!}\frac{2^{N-2M}}{\left(  N-M\right)
!}T^{M}R^{N-M}\int_{-\pi}^{+\pi}\frac{d\Lambda}{2\pi}\cos\left[  \left(
N_{\alpha}-N_{\beta}\right)  \Lambda\right]  \left[  \cos\Lambda\right]
^{N-M}\left[  2\cos\Lambda\right]  ^{M} \label{app-12}%
\end{equation}
or, when (\ref{app-8}) is inserted:%
\begin{equation}
\frac{N!}{m_{A}!m_{B}!}\frac{2^{-M}}{\left(  N-M\right)  !}T^{M}R^{N-M}
\label{app-13}%
\end{equation}
The summation over $m_{A}$ and $m_{B}$ with constant sum $M$ then gives:%
\begin{equation}
\frac{N!}{M!\left(  N-M\right)  !}T^{M}R^{N-M} \label{app-14}%
\end{equation}
which provides the probability of detecting $M$ particles, independently of
which of the 4 detectors is activated.\ This probability is maximal when:%
\begin{equation}
\frac{N-M}{M}\frac{T}{R}\sim1~\ \ \ \ \ \ \ \text{\ or \ \ \ \ \ \ \ }\frac
{M}{N}\sim T \label{app-14-bis}%
\end{equation}
The larger the transmission coefficient $T$, the larger the most likely value
of $M$, as one could expect physically.\ Finally, a summation of
(\ref{app-14}) over $M$ between $0$ and $N$ gives:%
\begin{equation}
\left(  R+T\right)  ^{N}=1 \label{app-15}%
\end{equation}
and the total probability is $1$, as expected.

\bigskip\bigskip

\begin{center}
APPENDIX II
\end{center}

\bigskip

In this appendix, we investigate how Eq. (\ref{19}) is changed when the
initial state $\left\vert \Phi_{0}\right\rangle $ is different from the double
Fock state considered in (\ref{1}).

(a) Coherent states

We first assume that each of the modes $\alpha$, $\beta$ is in a coherent
state with phases $\phi_{\alpha}$, $\phi_{\beta}$ and the same amplitude $E$:%
\begin{equation}
\left\vert \Phi_{0}\right\rangle =\left\vert \phi_{\alpha}\right\rangle
\otimes\left\vert \phi_{\beta}\right\rangle \label{aa-1}%
\end{equation}
with the usual expression of the coherent states:%
\begin{equation}
\left\vert \phi_{\alpha,\beta}\right\rangle \sim\sum_{r=0}^{\infty}%
\frac{\left[  E~e^{i\phi_{\alpha,\beta}}\right]  ^{r}}{\sqrt{r!}}\left\vert
N_{\alpha,\beta}=r\right\rangle \label{aa-2}%
\end{equation}
The calculation of \S \ \ref{PVR} is then simplified since this state is a
common eigenvector of both annihilation operators $a_{\alpha}$ and $a_{\beta}%
$. There is no need to introduce conservation rules, and neither $\lambda$ nor
$\Lambda$ enter the expressions.\ Eq. (\ref{19}) becomes:%
\begin{equation}%
\begin{array}
[c]{l}%
\mathcal{P(}m_{1},m_{2},m_{3},m_{4})\sim\left[  1+\cos\left(  \zeta
+\phi_{\alpha}-\phi_{\beta}\right)  \right]  ^{m_{1}}\left[  1-\cos\left(
\zeta+\phi_{\alpha}-\phi_{\beta}\right)  \right]  ^{m_{2}}\\
\multicolumn{1}{r}{\times\left[  1+\cos\left(  \theta+\phi_{\beta}%
-\phi_{\alpha}\right)  \right]  ^{m_{3}}\left[  1-\cos\left(  -\theta
+\phi_{\beta}-\phi_{\alpha}\right)  \right]  ^{m_{4}}}%
\end{array}
\label{aa-3}%
\end{equation}

Here, no $\lambda$ distribution occurs, in contrast with (\ref{19}): the
relative phase of the two states is perfectly defined and takes the exact
value $\phi_{\alpha}-\phi_{\beta}$. Now, we can also assume that the initial
phases of the two coherent states completely random.\ Then, an average over
all possible values of $\phi_{\alpha}$and $\phi_{\beta}$ leads to:%
\begin{equation}%
\begin{array}
[c]{l}%
\mathcal{P(}m_{1},m_{2},m_{3},m_{4})\sim\int\frac{d\phi}{2\pi}~\left[
1+\cos\left(  \zeta+\phi\right)  \right]  ^{m_{1}}\left[  1-\cos\left(
\zeta+\phi\right)  \right]  ^{m_{2}}\\
\multicolumn{1}{r}{\left[  1+\cos\left(  -\theta+\phi\right)  \right]
^{m_{3}}\left[  1-\cos\left(  -\theta+\phi\right)  \right]  ^{m_{4}}}%
\end{array}
\label{aa-4}%
\end{equation}
We now obtain an expression that is similar to Eq. (\ref{19}), but with a
difference: the terms of the product in the integral are always positive, as
if the quantum angle $\Lambda$ had been set equal to zero; no violation of
Bell inequalities is therefore possible. This was expected: with coherent
states, the phase pre-exists the measurement and is not created under the
effect of quantum measurement, as was the case with Fock states; an unknown
classical variable does not lead to violations local realism. Moreover, the
requirement of measuring all particles does not apply in this case, since the
initial state does not have an upper bound for the populations.

\bigskip

(b) Phase state

We now assume choose a state that has a fixed number of particles, but a well
defined phase $\phi_{0}$ between the two modes:%
\begin{equation}
\left\vert \Phi_{0},N\right\rangle =\frac{1}{\sqrt{N!}}\left[  e^{i\phi_{0}%
}a_{\alpha}^{\dagger}+a_{\beta}^{\dagger}\right]  ^{N}\left\vert
0\right\rangle =\sqrt{N!}\sum_{q=0}^{N}~\frac{e^{iq\phi_{0}}}{q!\left(
N-q\right)  !}~\left(  a_{\alpha}^{\dagger}\right)  ^{q}\left(  a_{\beta
}^{\dagger}\right)  ^{N-q}\left\vert 0\right\rangle \label{aa-5}%
\end{equation}
with $N$ even.\ We assume that all particles are measured: $\sum_{i}m_{i}=N$.
The probability amplitude we wish to calculate is:%

\begin{equation}
C_{m_{1}\cdots m_{4}}=\frac{1}{\sqrt{\prod_{j}m_{j}!}}\left\langle
0\right\vert \prod_{j=1}^{4}a_{j}^{m_{j}}\left\vert \Phi_{0},N\right\rangle
\label{aaa-1}%
\end{equation}
where the $a_{i}$ are defined in (\ref{awm}) and written more generically in
(\ref{ai}). We will use two different methods to do the calculation, first a
method based on the specific properties of phase states, and then a more
generic method extending the results of \S \ \ref{PVR}.

(i) The phase state $\left\vert \Phi_{0},N\right\rangle $ is by definition a
state where all bosons are created in one single state $\left[  e^{i\phi_{0}%
}\left\vert \alpha\right\rangle +\left\vert \beta\right\rangle \right]
/\sqrt{2}$, none in the orthogonal state $\left[  -e^{i\phi_{0}}\left\vert
\alpha\right\rangle +\left\vert \beta\right\rangle \right]  /\sqrt{2}%
$.\ Therefore the action of the two annihilation operators:%
\begin{equation}%
\begin{array}
[c]{l}%
a_{\phi_{0}}=\frac{e^{-i\phi_{0}}a_{\alpha}+a_{\beta}}{\sqrt{2}}\\
a_{\phi_{0+\pi}}=\frac{-e^{-i\phi_{0}}a_{\alpha}+a_{\beta}}{\sqrt{2}}%
\end{array}
\label{aaa-2}%
\end{equation}
is straightforward: the former transforms $\left\vert \Phi_{0},N\right\rangle
$ into $\left\vert \Phi_{0},N-1\right\rangle $, the latter gives zero.\ Now,
we can use (\ref{ai}) and (\ref{aaa-2}) to expand each $a_{i}$ as:%
\begin{equation}
a_{j}=v_{j\alpha}e^{i\phi_{0}}\frac{a_{\phi_{0}}-a_{\phi_{0+\pi}}}{\sqrt{2}%
}+v_{j\beta}\frac{a_{\phi_{0}}+a_{\phi_{0+\pi}}}{\sqrt{2}} \label{aaa-3}%
\end{equation}
where the action of $a_{\phi_{0+\pi}}$ gives zero.\ We conclude that:%
\begin{equation}
a_{j}\left\vert \Phi_{0},N\right\rangle =\sqrt{\frac{N}{2}}(e^{i\phi_{0}%
}v_{j\alpha}+v_{j\beta})\left\vert \Phi_{0},N-1\right\rangle \label{aaa-4}%
\end{equation}
in which case:%
\begin{equation}
C_{m_{1}\cdots m_{4}}=\sqrt{\frac{N!}{2^{N}\prod_{j}m_{j}!}}\prod_{j=1}%
^{4}(e^{i\phi_{0}}v_{j\alpha}+v_{j\beta})^{m_{j}} \label{PSProb}%
\end{equation}
The probability is then:%
\begin{equation}%
\begin{array}
[c]{l}%
\displaystyle\mathcal{P}(m_{1},m_{2},m_{3},m_{4})=\frac{N!}{\prod_{j}%
2^{N}m_{j}!}\prod_{j=1}^{4}\left[  (e^{-i\phi_{0}}v_{j\alpha}^{\ast}%
+v_{j\beta}^{\ast})(e^{i\phi_{0}}v_{j\alpha}+v_{j\beta})\right]  ^{m_{j}}\\
\multicolumn{1}{c}{\displaystyle=\frac{N!}{4^{N}~m_{1}!...m_{4}!}\left[
(1+\cos(\zeta+\phi_{0})\right]  ^{m_{1}}\left[  (1-\cos(\zeta+\phi
_{0})\right]  ^{m_{2}}}\\
\multicolumn{1}{r}{\displaystyle\times\left[  (1+\cos(\theta-\phi_{0})\right]
^{m_{3}}\left[  (1-\cos(\theta-\phi_{0})\right]  ^{m_{4}}}%
\end{array}
\label{aaa-5}%
\end{equation}
As in case (a), we have a state for which the quantum angle vanishes so that
no violation of the BCHSH inequalities can take place.

\bigskip

(ii) We can also do the calculation by a method that is similar to that of
\S \ \ref{PVR}.\ From (\ref{aa-5}) and (\ref{aaa-1}), we obtain:%
\[
C_{m_{1}\cdots m_{4}}=\frac{\sqrt{N!}}{\sqrt{\prod_{j}m_{j}!}}\sum_{q=0}%
^{N}~\frac{e^{iq\phi_{0}}}{q!\left(  N-q\right)  !}~\prod_{j=1}^{4}%
~\left\langle 0\right\vert a_{j}^{m_{j}}\left(  a_{\alpha}^{\dagger}\right)
^{q}\left(  a_{\beta}^{\dagger}\right)  ^{N-q}\left\vert 0\right\rangle
\]
The calculation is the same as in \S \ \ref{PVR}, with $N_{\alpha}$ replaced
by $q$ and $N_{\beta}$ by $N-q$; the prefactors of (\ref{1})\ and (\ref{aa-5})
combine to introduce a factor $\sqrt{N!/q!\left(  N-q\right)  !}$, and the
equivalent of (\ref{ampl}) is now:%
\begin{equation}
\sqrt{N!}\sum_{q=0}^{N}e^{iq\phi_{0}}\int_{-\pi}^{\pi}\frac{d\mu}{2\pi
}e^{^{i\left(  N-2q\right)  \mu}}\prod_{j=1}^{4}\left(  v_{j\alpha}e^{i\mu
}+v_{j\beta}e^{-i\mu}\right)  ^{m_{j}} \label{aa-ampl}%
\end{equation}
In the probability, a sum over $q$ and $q^{\prime}$ appears, including term
$q\neq q^{\prime}$ corresponding to non-diagonal probability terms between two
different states $\left\langle N_{\alpha}=q^{\prime}~;N_{\beta}=N-q^{\prime
}\right\vert $ and $\left\vert N_{\alpha}=q~;N_{\beta}=N-q\right\rangle
$.\ One finally obtains:%
\begin{equation}%
\begin{array}
[c]{l}%
\displaystyle\mathcal{P(}m_{1},m_{2},m_{3},m_{4})=\frac{N!}{m_{1}!m_{2}%
!m_{3}!m_{4}!}2^{-N}\int_{-\pi}^{\pi}\frac{d\lambda}{2\pi}\int_{-\pi}^{\pi
}\frac{d\Lambda}{2\pi}~G\left(  \lambda,\Lambda\right)  ~\left[  \cos
\Lambda+\cos\left(  \zeta+\lambda\right)  \right]  ^{m_{1}}\\
\displaystyle\times\left[  \cos\Lambda-\cos\left(  \zeta+\lambda\right)
\right]  ^{m_{2}}\left[  \cos\Lambda+\cos\left(  \theta-\lambda\right)
\right]  ^{m_{3}}\left[  \cos\Lambda-\cos\left(  \theta-\lambda\right)
\right]  ^{m_{4}}%
\end{array}
\label{aa-6}%
\end{equation}
with:%
\begin{equation}
G\left(  \lambda,\Lambda\right)  =\sum_{q,q^{\prime}=0}^{N}e^{i\left(
q-q^{\prime}\right)  \left(  \phi_{0}-\lambda\right)  }~e^{i\left(
N-q-q^{\prime}\right)  \Lambda} \label{aa-7}%
\end{equation}
The result is therefore similar to (\ref{19}), except for the presence of the
function $G\left(  \lambda,\Lambda\right)  $, which introduces a distribution
of the phase $\lambda$ and of the quantum angle $\Lambda$.

Equations (\ref{aa-6}) and (\ref{aa-7}) are equivalent to (\ref{aaa-5}),
although they do not contain the same distribution $G\left(  \lambda
,\Lambda\right)  $.\ Equation (\ref{aaa-5}) corresponds to an infinitely
narrow distribution, since it can be obtained by replacing in (\ref{aa-6})
$G\left(  \lambda,\Lambda\right)  $ by the product $\delta(\lambda-\phi_{0}%
)$~$\delta(\Lambda)$; by contrast, (\ref{aa-7}) defines a distribution with
finite width.\ This illustrates the fact that, in (\ref{aa-6}), different
distributions $G\left(  \lambda,\Lambda\right)  $ may lead to the same set of probabilities.

(c) General state

Consider finally the more general state $\left\vert \Phi_{0}\right\rangle $
combining two modes with a fixed total number of particles can be written as:%
\begin{equation}
\left\vert \Phi_{0}\right\rangle =\sum_{q=0}^{N}x_{q}~\left\vert N_{\alpha
}=q~;~N_{\beta}=N-q\right\rangle \label{aa-8}%
\end{equation}
where the complex coefficients $x_{q}$ are arbitrary.\ The calculation is
similar to that of \S (ii) above, but now one obtains (\ref{aa-6}) with a
different expression of $G(\lambda,\Lambda)$:%
\begin{equation}
G(\lambda,\Lambda)=\sum_{q,q^{\prime}}x_{q}x_{q^{\prime}}^{\ast}%
~e^{i(N-q-q^{\prime})}e^{-i\lambda(q-q^{\prime})}\sqrt{q^{\prime}%
!(N-q^{\prime})!q!(N-q)!} \label{aa-9}%
\end{equation}
Depending on the choice of the coefficients $x_{q}$, one can build states in
which the initial phase is well determined, as in \S \ (ii), or completely
indetermined as for Fock states; a similar conclusion holds for the quantum
angle $\Lambda$.

\end{document}